\documentclass[useAMS,usenatbib]{mn2e}

\usepackage{amssymb}
\usepackage{amsmath}
\usepackage{graphicx}
\usepackage{epstopdf}
\usepackage{verbatim}
\usepackage{latexsym}
\usepackage{theorem}
\usepackage{fixltx2e}
\usepackage{lscape}

\title[Evolution and fate of very massive stars]{Evolution and fate of very massive stars}
\author[N. Yusof et. al]
{Norhasliza Yusof$^{1,2,3}$\thanks{E-mail: norhaslizay@um.edu.my}, Raphael
Hirschi$^{3,4}$, Georges Meynet$^{5}$, Paul A. Crowther$^{6}$,
\newauthor Sylvia Ekstr\"om$^{5}$, Urs Frischknecht$^{3,8}$, Cyril Georgy$^{3,7}$, Hasan Abu Kassim$^{1,2,10}$,  
\newauthor Olivier Schnurr$^{9}$
\\
$^{1}$Department of Physics, Faculty of Science, University of Malaya, 50603 Kuala Lumpur, Malaysia\\
$^{2}$Quantum Science Center, Faculty of Science, University of Malaya, 50603 Kuala Lumpur, Malaysia\\
$^{3}$Astrophysics, Lennard-Jones Laboratories, EPSAM, Keele University, ST5 5BG, Staffordshire, UK\\
$^{4}$Kavli Institute for the Physics and Mathematics of the Universe,
University of Tokyo, 5-1-5 Kashiwanoha, Kashiwa, 277-8583, Japan\\
$^{5}$Geneva Observatory, Geneva University, 1290 Sauverny,
Switzerland \\
$^{6}$ Department of Physics and Astronomy, University of
Sheffield, Hicks Building, Hounsfield Road, Sheffield S3 7RH\\
$^{7}$Centre de Recherche Astrophysique de Lyon, Ecole
Normale Sup\'{e}rieure de Lyon, 46, all\'{e}e d'Italie, F-69384 Lyon cedex 07, France\\
$^{8}$Dept. of Physics, University of Basel, Klingelbergstr. 82,
4056, Basel, Switzerland\\
$^{9}$Leibniz-Institut für Astrophysik Potsdam (AIP), An der Sternwarte 16, 14482 Potsdam, Germany\\
$^{10}$Institute of Space Science, Universiti Kebangsaan Malaysia, 43600 Bangi, Selangor, Malaysia}
\begin{document}


\pagerange{\pageref{firstpage}--\pageref{lastpage}} \pubyear{2013}

\maketitle

\label{firstpage}

\begin{abstract}
There is observational evidence that supports the existence of Very Massive Stars (VMS) in the local universe.
First, very massive stars ($M_{\textrm{ini}} \lesssim320\,M_\odot$) have been observed in the Large Magellanic Clouds (LMC). Second, there are observed SNe that bear the characteristics of Pair Creation Supernovae (PCSNe , also referred to as pair-instability SN) which have very massive stars as progenitors. The most promising candidate to date
is SN2007bi. 
In  order to investigate the evolution and fate of nearby very massive stars, we calculated a new grid of models for such objects, for solar, LMC and SMC metallicities, which covers the initial mass range from 120 to 500 $M_\odot$. Both rotating and non-rotating models were calculated using the Geneva stellar evolution code and evolved until at least the end of helium burning and for most models until oxygen burning. 
Since very massive stars have very large convective cores during the Main-Sequence phase, their evolution is not so much affected by rotational mixing, but more by
mass loss through stellar winds. Their evolution is never far from a homogeneous evolution even without rotational mixing. All the VMS, at all the metallicities studied here, end their life as WC(WO) type Wolf-Rayet stars.  Due to very important mass losses through stellar winds, these stars may have luminosities during the advanced phases of their evolution similar to stars with initial masses between 60 and 120 $M_\odot$. A distinctive feature which may be used
to disentangle Wolf-Rayet stars originating from VMS from those originating from lower initial masses would be the enhanced abundances of Ne and Mg at the surface of WC stars. This feature is however not always apparent depending on the history of mass loss. At solar metallicity, none of our models is expected to explode as a PCSN. At the metallicity of the LMC, only stars more massive than 300 $M_\odot$ 
are expected to explode as PCSNe.
At the SMC metallicity, the mass range for the PCSN progenitors is much larger and comprises stars with initial masses between about 100 and 290 $M_\odot$.
All VMS stars in the metallicity range studied here produce either a type Ib or a type Ic SN but not a type II SN.
We estimate that the progenitor of SN2007bi, assuming a SMC metallicity, had an initial mass between 160 and 175 $M_\odot$.
None of  models presented in this grid produce GRBs or magnetars. 
They lose too much angular momentum by mass loss or avoid the formation of a BH by producing a completely disruptive PCSN.
\end{abstract}


\section{Introduction}
In the present work, we call Very Massive Stars
(VMS), stars with initial masses superior to
100 $M_\odot$. For a long time, the evolution of such
stars were considered only in the framework of
Pop III stars. Indeed, it was expected that, only in
metal free environments, could such massive stars
be formed, since the absence of dust, an efficient
cooling agent, would prevent a strong
fragmentation of the proto-stellar cloud \citep{BV99, Abel02}\footnote{Note, however, that the most recent star formation simulations find lower-mass stars forming in groups, similarly to present-day star formation \citep{Stacy10, Greif10}}. 
It came therefore as a surprise when it was
discovered that the most metal poor low mass
stars, likely formed from a mixture between the
ejecta of these Pop III stars and pristine
interstellar medium, did not show any signature
of the peculiar nucleosynthesis of the VMS \citep{HEGER02,UN02,CH02, 2005Natur}. While
such observations cannot rule out the existence of
these VMS in Pop III generations (their
nucleosynthetic signature may have been erased by
the more important impact of stars in other mass
ranges), it seriously questions the
importance of such object for understanding the
early chemical evolution of galaxies. Ironically, when the importance of VMS in the context of the first
stellar generations fades, they appear as
potentially interesting objects in the framework of
present day stellar populations. 

For a long time, observations favored a present-day upper mass limit for stars
around 150 $M_\odot$ \citep{FI05,OC05}.
Recently, however, \citet{PAC10} have re-assessed the properties of the brightest
members of the R136a star cluster, revealing exceptionally high
luminosities. The comparison between main sequence evolutionary models for rotating and
non-rotating stars and observed spectra resulted in high current ($\leq$ 265
$M_{\odot}$) and initial ($\leq  $ 320 $M_{\odot}$) masses for
these stars.

In addition, the advent of all sky transient surveys --
unbiased towards specific nearby massive star forming galaxies -- has
produced a population of super-luminous supernovae \citep[SLSNe,][]{GALYAM12}, 
which may have VMS as progenitors.
%
%
SN2006gy was the first example of such a supernova for which a pair-creation
supernova (PCSN) was
suspected \citep{SN07}, while still stronger evidence was
reported by \citet{GALYAM09} for SN2007bi from a metal-poor dwarf galaxy
at z=0.128. They derived a core helium mass of 100 $M_{\odot}$ for
SN2007bi, and estimated an initial mass of 200 $M_{\odot}$ for its
progenitor, although substantially higher initial masses were inferred by
\citet{YH11}. Most recently, other SN2007bi-like examples have been
identified: PTF10 nmn (Gal-Yam, submitted; Yaron, in
preparation) and PS1-11ap (Kotak et al., in preparation), suggesting that
VMS and PCSN occur at the current epoch.

The above observations trigger a new interest in the formation, evolution and fate of very massive stars and
in particular stimulate the present study.
Our aim is mainly  twofold: 1) to provide grids for VMS at three different metallicities
with and without rotation. These grids are useful to interpret observations of very 
luminous objects; 2) to study the evolutionary scenario and the final fate of such objects. We want to address questions such as
the kind of stars and of supernova explosion (if any) that VMS produce.

In order to realise these two aims, we calculated a new grid of VMS models at solar ($Z=0.014$), Large Magellanic Cloud (LMC, $Z=0.006$) and Small Magellanic Cloud (SMC, $Z=0.002$) metallicities using the Geneva stellar evolution code \citep{GVAcode} including the modifications implemented to follow the advanced stages as described in \citet{psn04}.
This work extends the main sequence models for VMS presented in
\cite{PAC10} and complements the low and high mass stellar grids at solar
metallicity ($Z=0.014$) calculated by \cite{SE12} and \cite{GC12} as well as their extension to lower metallicities (Georgy et al. in prep., Eggenberger et al. in prep.). 

In Sect.~2, we present the physical ingredients of the models. The description of the results is given in Sect.~3. The Wolf-Rayet stars originating from VMS are
discussed in Sect.~4. The final fates of the VMS is the subject of Sect.~5. Conclusions and perspectives are given in Sect.~6.

\section[]{Physical ingredients of the models}
The various input physics parameters are the same as the ones used in the new grids of rotating models published in \citet{SE12}, making this grid of VMS consistent with their published grid. We just recall here a few important points:
\begin{itemize}
 \item The initial abundances for the models are listed in Table \ref{ini}. The mixture of heavy elements ($Z$) is taken from \cite{APS05} except for the Ne abundances adopted from \cite{Cunha06} and the isotopic ratios are taken from \citet{Lo03}. 
\item Reaction rates are taken mostly from NACRE \citep{AN99}. The full list of updated rates as well as a short description of the effects on stellar evolution are presented in \cite{SE12}.
\item Neutrino energy loss in plasma, pair and photo-neutrino process is taken from \cite{Itoh89} and \cite{Itoh96}.
\item Opacities is taken from OPAL \citep{IR96} and completed with low temperature opacities from \citet{AF05} adapted for the high Ne abundance.
\item For convection, the Schwarzschild criterion is used and a modest overshooting with an overshoot parameter,
$\mathrm{d_{over}}/H_P$ = 0.10, is used for core hydrogen and helium burning only.
\item The outer convective zone is treated according to the mixing length theory, using $\alpha_\mathrm{MLT}$ = 1.0. As explained in \citet[][and references therein]{SE12},
$\alpha_\mathrm{MLT}=l/H_\rho$, where $l$ is the mixing length and $H_\rho$ the density scale height to avoid an unphysical density inversion in the envelope.
\item The treatment of rotation in the Geneva code has been described in \citet{MMr10}. For horizontal turbulence, we used the diffusion coefficient from \citet{Za92} and for the secular shear turbulence, we used the diffusion coefficient of \citet{ROTII}.
\item Note that the effects due to the creation of electron-positron pairs is not included in the equation of state but this does not affect the conclusions of this paper as discussed below.
\end{itemize}

\begin{table}
\centering
\caption{Hydrogen ($^1{\rm H}$), helium isotopes ($^3{\rm He}$, $^4{\rm He}$) and metal (Z) mass fractions for the chemical abundances in our models.}\label{ini}
\begin{tabular}{cccccc}
\hline
      &$^1{\rm H}$  &$^3{\rm He}$ &$^4{\rm He}$ & Z \\
\hline
Solar &0.7200   &4.414e-5 &0.2659 &0.014 \\
LMC   &0.7381   &4.247e-5 &0.2559 &0.006 \\
SMC   &0.7471   &4.247e-5 &0.2508 &0.002 \\
\hline
\end{tabular}
 \end{table}

\subsection{Mass loss}\label{dMdt}

Mass loss strongly affects the evolution of very massive stars as we shall describe below. It is therefore important to understand the different mass loss prescriptions used and how they relate to each other. In this study, we used the following prescriptions. For main-sequence stars, we used the prescriptions for radiative line driven winds from \citet{VN01}, which compare rather well with observations \citep{PAC10,muijres11}. 
For stars in a domain not covered by the Vink et~al prescription, we applied the \citet{deJager88} prescription to models with $\log (T_\text{eff}) > 3.7$. For $\log (T_\text{eff}) \leq 3.7$, we performed a linear fit to the data from \citet{sylvester98} and \citet{vloon99} \citep[see][]{Crowther01}. The formula used is given in Eq.\,2.1 in \citet{BHP12}.

In the calculations, we consider that a star becomes a WR when the surface hydrogen mass fraction, $X_s$, becomes inferior to 0.3 and the effective temperature, $\log(T_\text{eff})$, is greater than 4.0. 
The mass loss rate used during the WR phase depends on the WR sub-type. For the eWNL phase (when $0.3>X_s>0.05$), the 
\citet{GH08} recipe is used (in the validity domain of this prescription, which usually covers most of the eWNL phase). In many cases, the WR mass-loss rate of \citet{GH08} is lower than the rate of \citet{VN01}, in which case, we used the latter. For the eWNE phase -- when $0.05>X_s$ and the ratio of the mass fractions of ($^{12}$C$+\,^{16}$O)/$^4$He$< 0.03$ -- and WC/WO phases -- when ($^{12}$C$+\,^{16}$O)/$^4$He$ > 0.03$-- we used the corresponding prescriptions of \citet{NL00}. 
Note also that both the \citet{NL00} and \citet{GH08} mass-loss rates account for clumping effects \citep{muijres11}.

As is discussed below, the mass loss rates from \citet{NL00} for the eWNE phase are much larger than in other phases and thus the largest mass loss occurs during this phase. In \citet{PAC10}, the mass loss prescription from \citet{NL00} was used for both the eWNL and eWNE phases. The current models thus lose less mass than those presented in \citet{PAC10} during the eWNL phase. 

The metallicity dependence of the mass loss rates is included in the following way. The mass loss rate used at a given metallicity, $\dot{M}(Z)$, is the mass loss rate at solar metallicity, $\dot{M}(Z_\odot)$, multiplied by the ratio of the metallicities to the power of $\alpha$: $\dot{M}(Z)= \dot{M}(Z_\odot)(Z/Z_\odot)^\alpha$.
$\alpha$ was set to 0.85 for the O-type phase and WN phase and 0.66 for the WC and WO phases; and
for WR stars the initial metallicity rather than the actual surface metallicity was used in the equation above following \citet{EV06}.
$\alpha$ was set to 0.5 for the \citet{deJager88} prescription. Finally, $\alpha$ was set to 0 (no dependence) if $\log (T_\text{eff}) \leq 3.7$ (note that none of the models presented in this study reach such low effective temperatures).

For rotating models, we applied to the radiative mass-loss rate the correction factor described in \citet{ROTVI}
\begin{eqnarray}
\dot{M}(\Omega) &=& F_{\Omega}\cdot \dot{M}(\Omega=0)  \nonumber \\
& &\text{with} \hspace{.3cm}F_{\Omega}=\frac{(1-\Gamma)^{\frac{1}{\alpha}-1}}{\left[ 1-\frac{\Omega^2}{2\pi G \rho_\text{m}} - \Gamma \right]^{\frac{1}{\alpha}-1}},
\label{EqMdotRot}
\end{eqnarray}
where $\Gamma=L/L_\text{Edd}=\kappa L / (4\pi cGM)$ is the Eddington factor (with $\kappa$ the total opacity), and $\alpha$ the $T_\text{eff}-$dependent force multiplier parameter.

\subsection{Models computed}\label{models}

\begin{table}
\caption{Main properties at the start of the core H-burning phase (ZAMS). {$M_{\rm ini}$ is the initial mass, $Z_{\rm ini}$ is
the initial metallicity, $v_\mathrm{eq}$ is the equatorial velocity in $\rm{km}$ $\rm{s}^{-1}$, $L_{\rm ZAMS}$ is the luminosity, $T_\mathrm{eff}^{\rm ZAMS}$ is
the effective luminosity and $\Gamma_\mathrm{Edd}^{\rm ZAMS}$ is the Eddington limit.} }\label{Table:startH}
\begin{tabular}{lccccr}
\hline
 $M_{\rm ini}$  &$Z_{\rm ini}$  &${v_\mathrm{eq}}$ &$L_{\rm ZAMS}$ &$T_\mathrm{eff}^{\rm ZAMS}$  &$\Gamma_\mathrm{Edd}^{\rm ZAMS}$    \\
\hline
 120     &0.014   &0   &6.231  &4.729  &0.376   \\
 150     &0.014   &0   &6.383  &4.736  &0.426   \\
 200     &0.014   &0   &6.567  &4.739  &0.489   \\
 300     &0.014   &0   &6.812  &4.734  &0.572   \\
 500     &0.014   &0   &7.099  &4.705  &0.666   \\
 \hline
 120     &0.014   &373   &6.230  &4.727  &0.374   \\
 150     &0.014   &395   &6.373  &4.723  &0.417   \\
 200     &0.014   &411   &6.558  &4.723  &0.479   \\
 300     &0.014   &426   &6.804  &4.716  &0.564   \\
 500     &0.014   &361   &7.098  &4.694  &0.717   \\
 \hline
 120    &0.006    &0   &6.227  &4.750 &0.376  \\
 150    &0.006    &0   &6.379  &4.758 &0.426 \\
 500    &0.006    &0   &7.091  &4.754 &0.660 \\
\hline
 120     &0.006   &400  &6.214 &4.737  &0.365  \\
 150     &0.006   &404  &6.377 &4.756  &0.425  \\
 200     &0.006   &440  &6.552 &4.751  &0.477  \\
 300     &0.006   &468  &6.798 &4.753  &0.560  \\
 500     &0.006   &415  &7.091 &4.754  &0.666  \\
\hline
 150     &0.002   &436  &6.377  &4.778  &0.426  \\
 200     &0.002   &443  &6.550  &4.776  &0.477  \\
 300     &0.002   &509  &6.793  &4.780  &0.557 \\
 \hline
\end{tabular}
\end{table}

We calculated models of 120, 150, 200, 300 and 500 $M_\odot$.
The initial properties of the models are listed in
Table~\ref{Table:startH}. The ZAMS is chosen as the time when $0.3\%$ in mass fraction of hydrogen is burnt. 
Although we focused our study on rotating stellar
models, we also calculated non-rotating models at solar and
LMC metallicities. 
{The initial surface equatorial velocity was chosen in order to have the same ratio of the initial to critical velocity, $\upsilon_{\rm ini}/\upsilon_{\rm crit}$, of 0.4 on the ZAMS for all models in line with \citet{SE12}. The surface velocity corresponding to this ratio increases with initial mass, except for the 500\,$M_\odot$, for which the very high luminosity reduces the critical velocity \citep[see][for more details]{MM00}.} This choice of initial velocities corresponds to an average velocity on the main sequence of 97 km s$^{-1}$ for the 120 $M_\odot$ stellar model and 141 km s$^{-1}$ for 500 $M_\odot$ model.

\section{Results of VMS models}
\subsection{VMS evolve nearly homogeneously}

\begin{figure}
\centering
\includegraphics[width=0.5\textwidth,clip=]{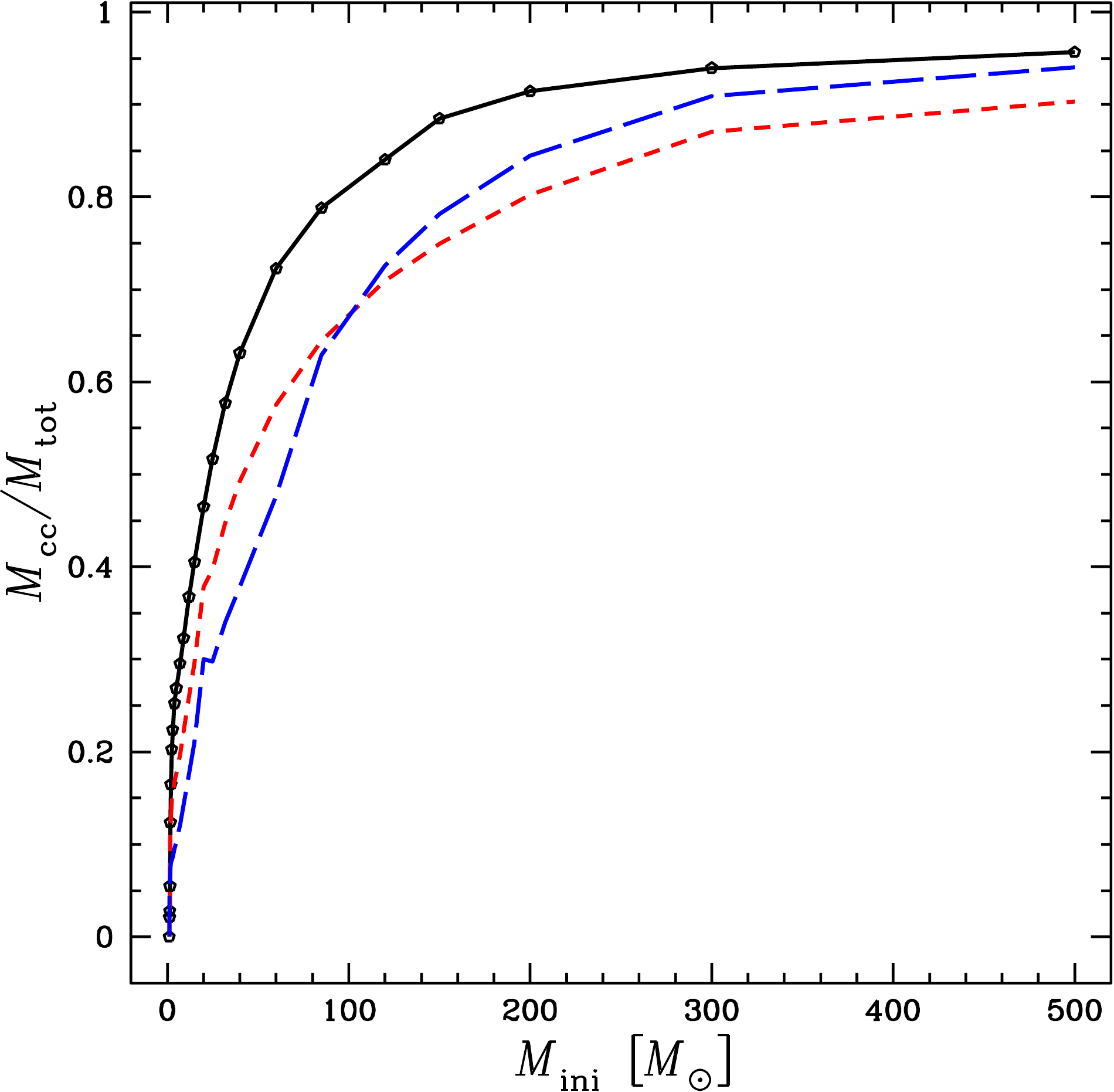}
\caption{Mass fraction of the convective core in non-rotating
solar metallicity models. The models with initial masses superior
or equal to 150 $M_\odot$ are from the present work. Models
for lower initial masses are from Ekstr\"om et al. (2012).
The continuous line corresponds to the ZAMS, the short-dashed line to
models when the mass fraction of hydrogen at the centre, $X_c$,  is 0.35, and
the long-dashed line to models when $X_c$ is equal to 0.05.
}\label{fig:ccom}
\end{figure}

\begin{figure*}
\centering
\begin{tabular}{cc}
\includegraphics[width=0.4\textwidth,clip=]{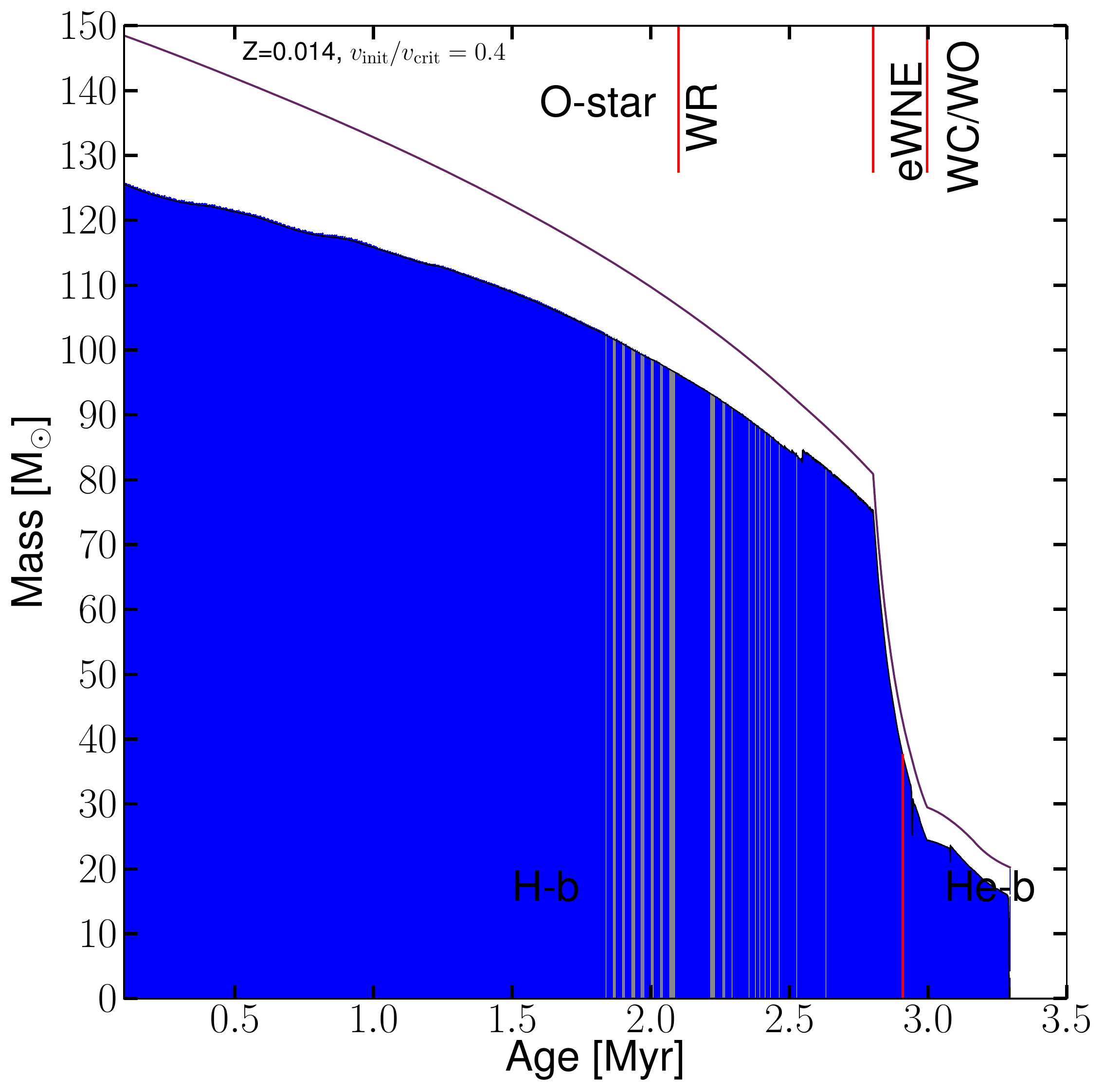} &
\includegraphics[width=0.4\textwidth,clip=]{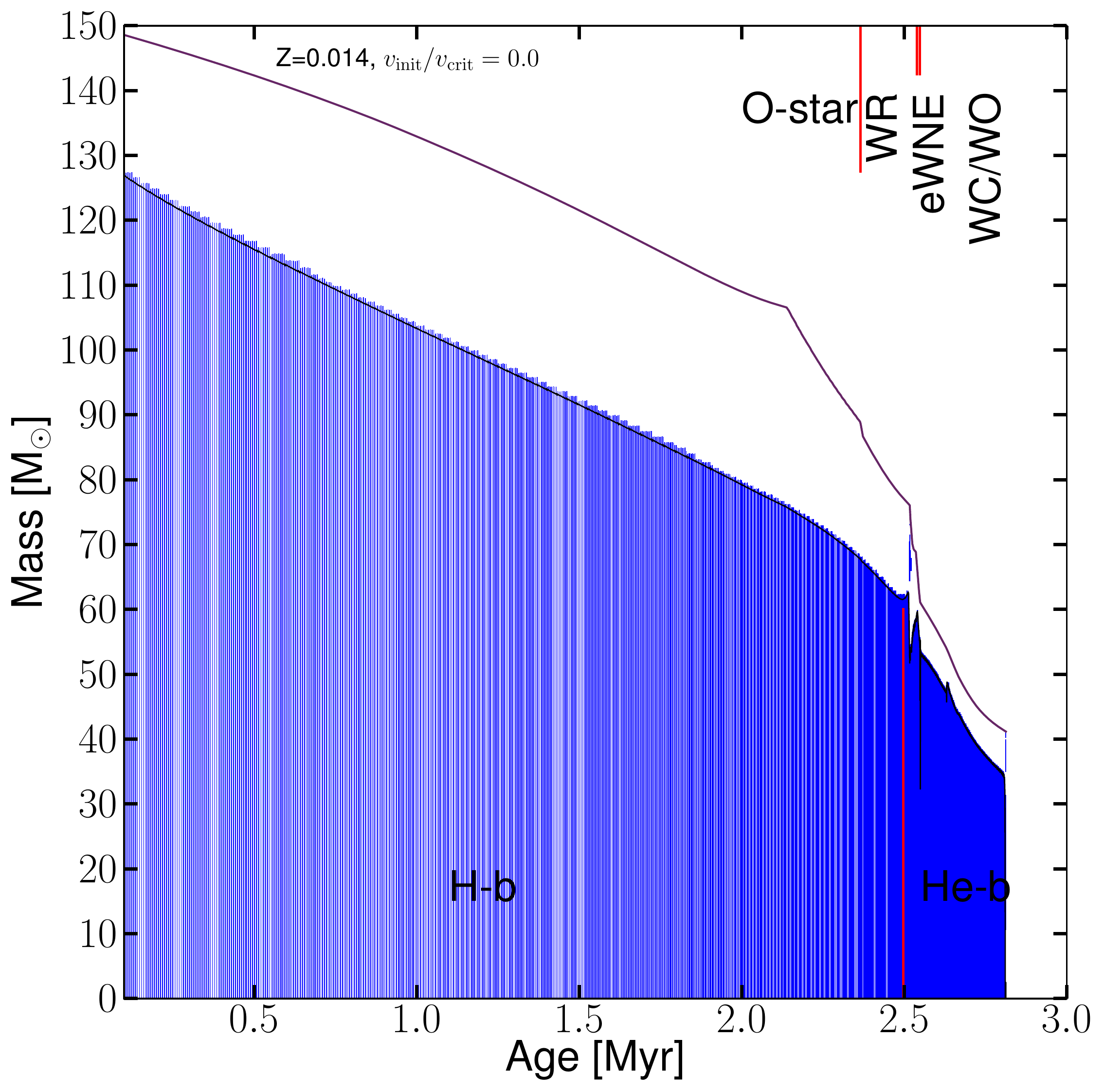} \\
\includegraphics[width=0.4\textwidth,clip=]{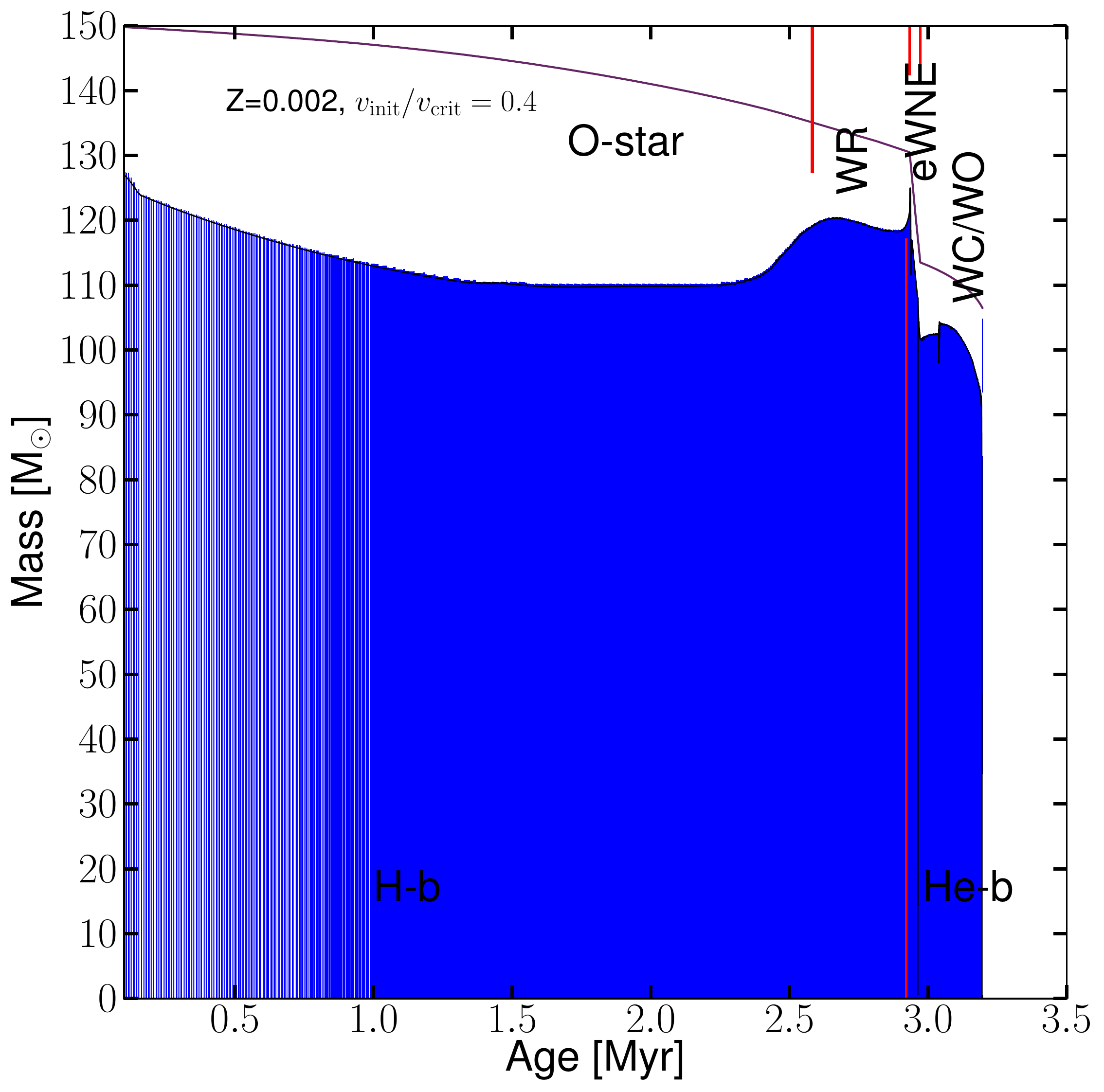} &
\includegraphics[width=0.4\textwidth,clip=]{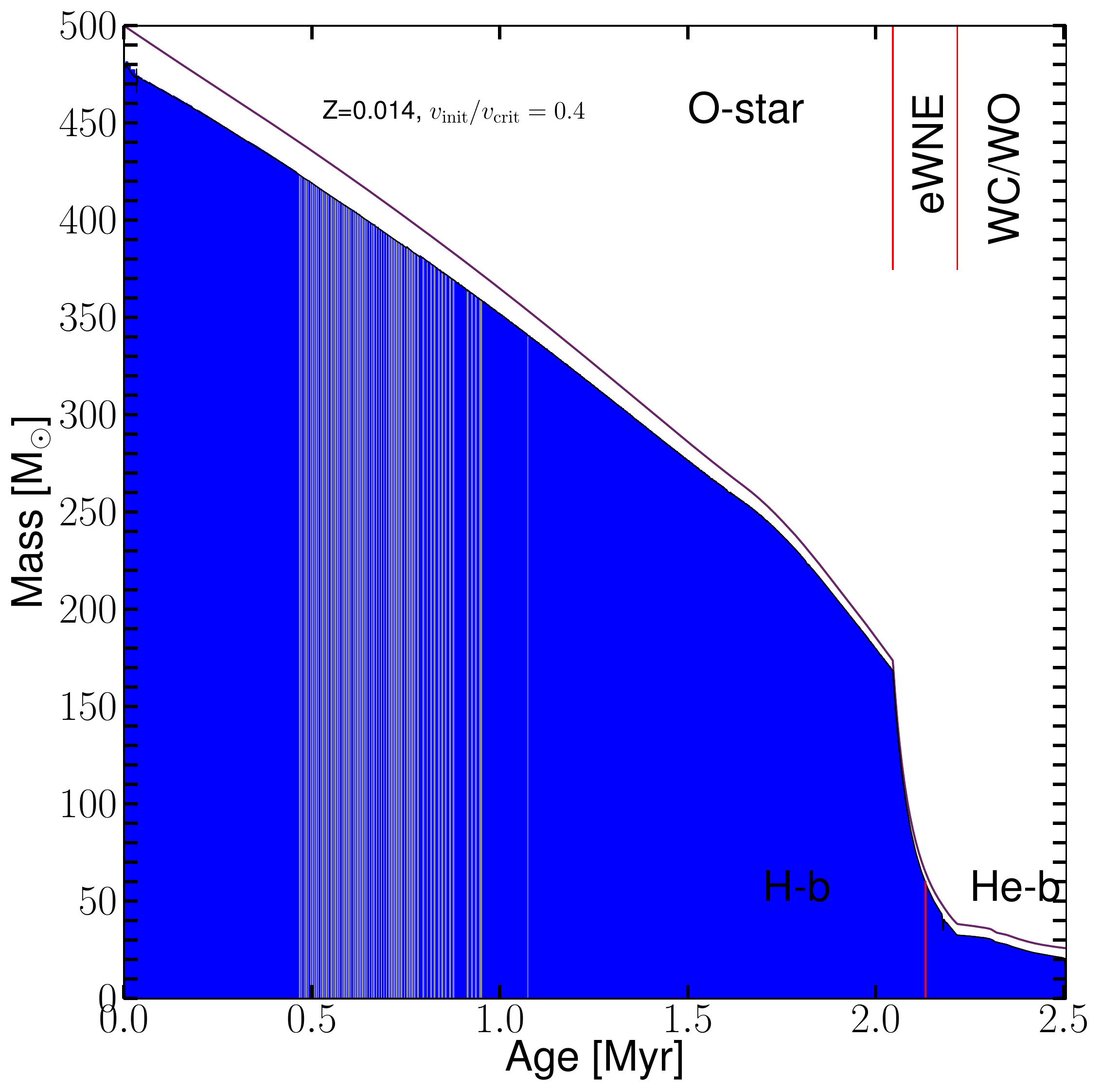} 
\end{tabular}
\caption{Structure evolution as a function of age for selected models: solar metallicity 150 $M_\odot$ rotating ({\it top-left}) and non-rotating ({\it top-right}) models, rotating SMC metallicity 150 $M_\odot$ model ({\it bottom-left}) and rotating solar metallicity 500 $M_\odot$ model ({\it bottom-right}). The blue zones represent the convective regions. The top solid black line indicates the total mass of the star and vertical red markers are given for the different phases (O-type, WR=eWNL, eWNE and WC/WO) at the top of the plots. The transition between H- and He-burning phases is indicated by the red vertical line at the bottom of the plots.}\label{fig:kip_age}
\end{figure*}

Probably the main characteristic that makes VMS quite different from their lower mass siblings is the fact that
they possess very large convective cores during the MS phase. 
To illustrate this last point, Fig.~\ref{fig:ccom} shows the convective core mass fraction for 
non-rotating massive stars at solar metallicity.
It is apparent that the convective cores for masses above 150 $M_\odot$
extend over more than 75\% of the total mass of the star.

Fig.~\ref{fig:kip_age} shows how age, metallicity and rotation influence this mass fraction. 
Comparing the {\it top-left} and {\it bottom-left} panels showing the rotating 150 $M_\odot$ models at solar and SMC metallicities ($Z$), respectively, we can see that the convective core occupies a { very slightly larger fraction of the total mass at SMC metallicity on the ZAMS. As for lower-mass massive stars, this is due to a lower CNO content leading to higher central temperature. This effect is counterbalanced by the lower opacity (especially at very low metallicities) and the net change in convective core size is small. As the evolution proceeds mass loss is weaker at lower $Z$ and thus the total mass decreases slower than the convective core mass. This generally leads to a smaller fraction of the total mass occupied by the convective core in the SMC models.}

We can see the impact of rotation by comparing the rotating ({\it top-left}) and non-rotating ({\it top-right}) 150 $M_\odot$ models. The convective core size remains higher in the rotating model due to the additional mixing in radiative zones. We can see that rotation induced mixing can even lead to an increase of the convective core size as is the case for the SMC model ({\it bottom-left}). This increase is typical of quasi-homogeneous evolution also found in previous studies \citep[see][and citations therein]{YDL12}.
The rotating 500 $M_\odot$ model ({\it bottom-right} panel) evolves quasi-homogeneously throughout its entire evolution, even with an initial ratio of the velocity to the critical velocity of 0.4. 

This feature of the most massive stars 
is a key factor governing their evolution as is discussed below.

\subsection{Evolutionary tracks and lifetimes}

\begin{figure}
\center
\includegraphics[width=0.5\textwidth,clip=]{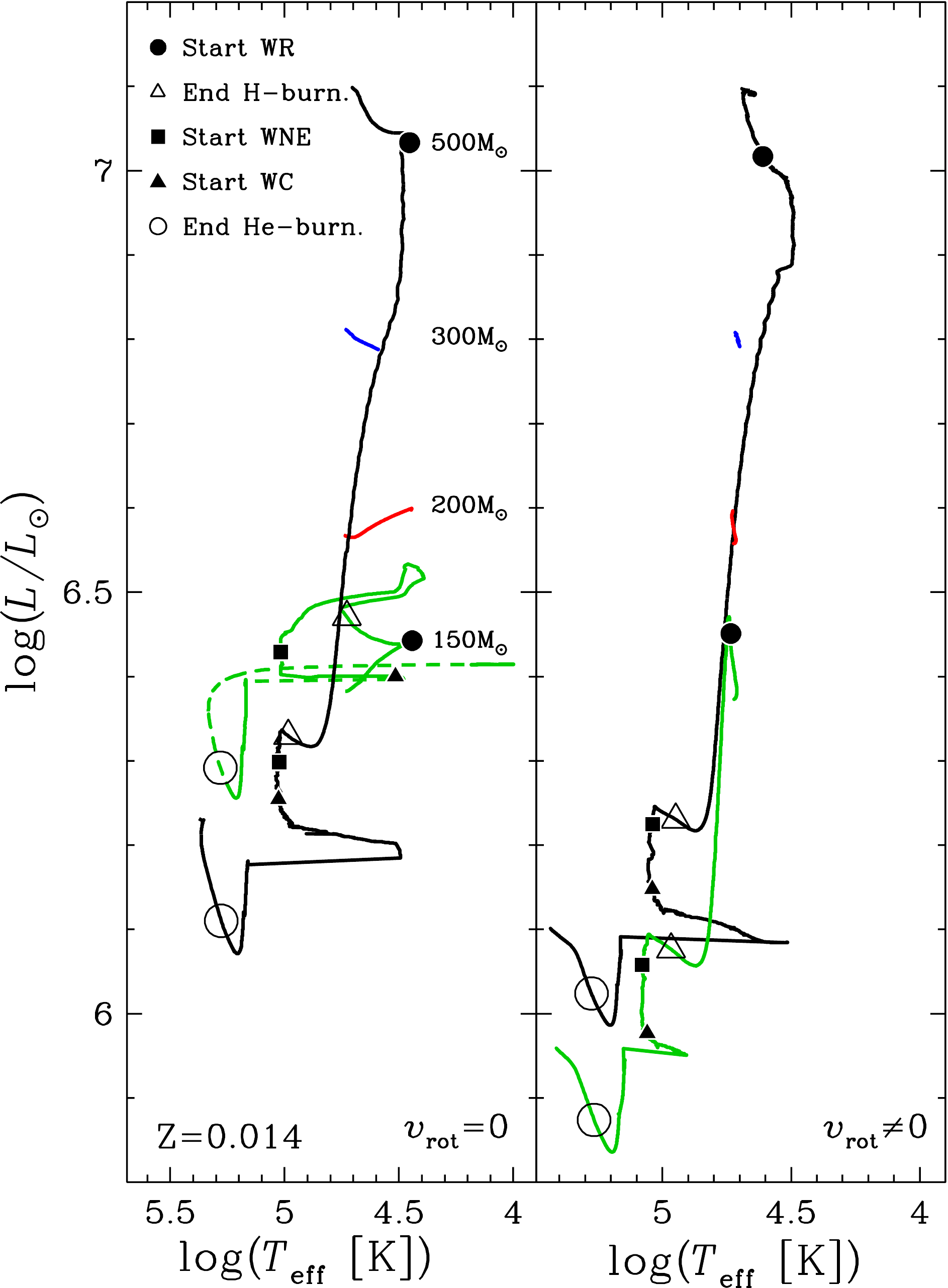}
\caption{HR diagram from 150 up to 500 $M_\odot$ at solar 
metallicity for non-rotating ({\it left}) and rotating ({\it right}) models, respectively. Key stages are indicated along the tracks. Only the first portion (up to start of WR phase) of the tracks for the 200 and 300 $M_\odot$ are shown.}\label{fig:hrdsolar}
\end{figure}

\begin{figure}
\center
\includegraphics[width=0.5\textwidth,clip=]{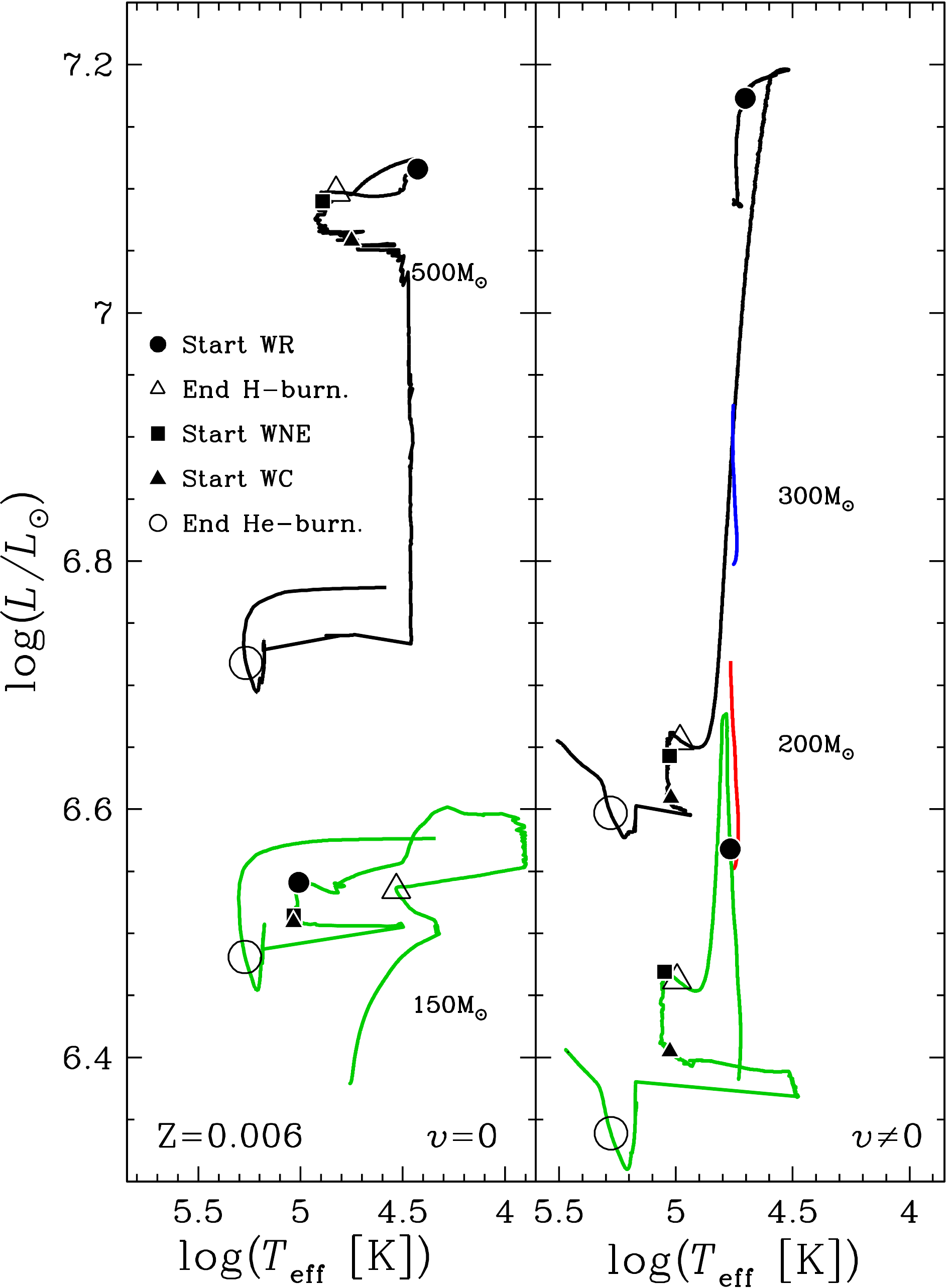}
\caption{Same as Fig.~\ref{fig:hrdsolar} for LMC models ($Z=0.006$).}
\label{fig:hrdlmc}
\end{figure}

\begin{figure}
\center
\includegraphics[width=0.5\textwidth,clip=]{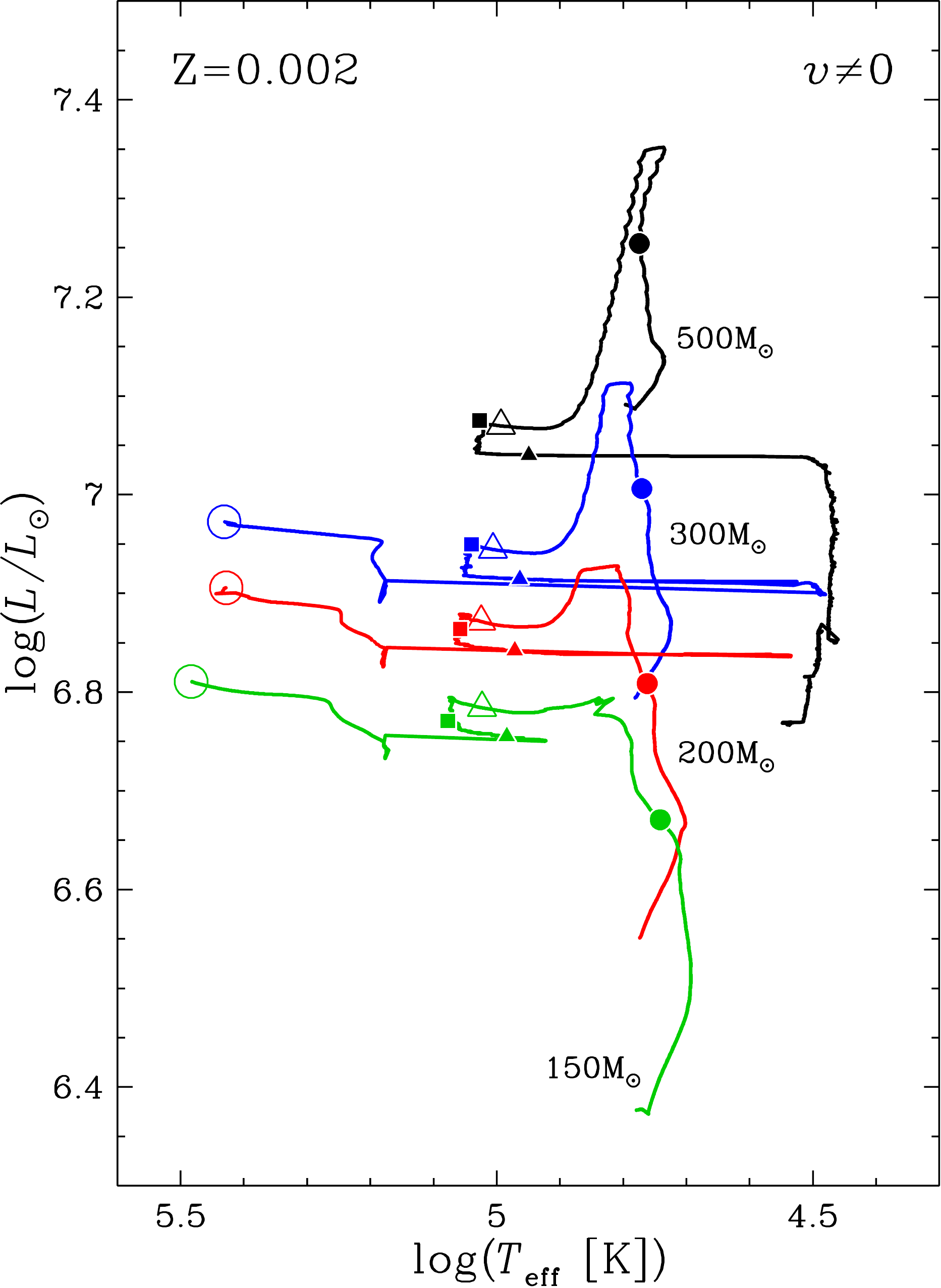}
\caption{Same as Fig.~\ref{fig:hrdsolar} ({\it right}) for SMC rotating models ($Z=0.002$).}
\label{fig:hrdsmc}
\end{figure}

In Figs.~\ref{fig:hrdsolar}, \ref{fig:hrdlmc} and \ref{fig:hrdsmc}, we present the evolutionary tracks of our present models with initial masses
between 150 and 500 $M_\odot$ for the three metallicities considered.
Other properties of the present stellar models are given at various evolutionary stages in Tables~\ref{Table:endH}, \ref{Table:endHe} and
\ref{Table:endmod}.


A very first striking feature is that these massive star evolve vertically in the HR diagram (HRD) covering only very restricted ranges in effective temperatures but a 
very large range in luminosities. This is typical of an evolution mainly governed by mass loss and also by a strong internal mixing (here due to convection).

Let us now describe in more details the evolution of the non-rotating 500 $M_\odot$ model at solar metallicity (see Fig.~\ref{fig:hrdsolar}).
In general, the luminosity of stars increases during the MS phase. Here we have that during that phase, 
the luminosity decreases slightly by about 0.1 dex.
This is the consequence of very high mass loss rates (of the order of $7 \times 10^{-5}$ $M_\odot$ per year) already at very early evolutionary stages.

At an age of 1.43  million years, the mass fraction of hydrogen at the surface becomes inferior to 0.3, the star enters into the WR phase and has an actual mass decreased by about 40\%
with respect to its initial value. At that time the mass fraction of hydrogen in the core is 0.24. Thus this star enters the WR phase while still burning its hydrogen in its core and having nearly the same amount of hydrogen at the centre and at the surface, illustrating the nearly homogeneous nature of its evolution.  Typically for this model, the convective core encompasses nearly 96\% of the total mass on the ZAMS (see also Fig. \ref{fig:ccom}).

At an age equal to 2.00 Myr, the mass fraction of hydrogen is zero in the core ($X_c=0$). The star has lost a huge amount of mass through stellar winds and has at this stage an actual mass of 55.7 $M_\odot$.  So, since the entrance into the WR phase, the star has lost about 245 $M_\odot$, i.e. about half of its total mass. This strong mass loss episode translates into the HR diagram
by a very important decrease in luminosity. Note that when $X_c$ is zero, the convective core still encompass 80\% of the total stellar mass!

The core helium burning phase last for about 0.3 Myr, that means slightly more than 15\% of the MS lifetime. 
At the end of the core He burning phase, the actual mass of the star is 29.82 $M_\odot$, its age is 2.32 My, the mass fraction of helium at the surface is 0.26.
The last model has an age of 2.32 My, an actual mass of 29.75 and a helium surface abundance in mass fraction of 0.06.
The total WR phase lasts for 0.88 My, that means about 38\% of the total stellar lifetime.

It is interesting to compare the evolution of the 500 $M_\odot$ stellar model with that of the 150 $M_\odot$ model. In contrast to the 500 $M_\odot$ model, the 150 $M_\odot$ increases in luminosity during the MS phase. Looking at the HRD we see that the O-type star phases of the 150 and 500 $M_\odot$ models cover more or less the same effective temperature range. 
This illustrates the well known fact that the colors of stars for this mass range does not change much with the initial mass. 

When the stars enters into the WR phase, in contrast to the case of the 500 $M_\odot$ where the luminosity decreases steeply, the luminosity of the 150 $M_\odot$ model continues to increase a little. The luminosities of the two models when the hydrogen mass fraction at the surface becomes inferior to 10$^{-5}$ differ by just a little more than 0.1 dex.  The effective temperatures are similar. Thus one expects stars from very different initial masses to occupy similar positions in the HRD (the 500 $M_\odot$ star being slightly less luminous than the 150 $M_\odot$ during the WR phase).
We note that after the end of the core He-burning phase, the star evolves to the red and terminate its lifetime around an effective temperature of Log $T_{\rm eff}$ equal to 4. This comes from the  
core contraction at the end of core He-burning which releases energy and leads to an envelope expansion akin to the expansion of the envelope at the end of the MS \citep[see also][]{YH11}. 

The duration of the core H-burning phase of the 150 $M_\odot$ model is not much different from the one of the 500 $M_\odot$ model being 2.5 My instead of 2 My. 
The core He-burning lifetime lasts for 0.3 My as for the 500 $M_\odot$. 
The total duration of the WR phase is 0.45 My, thus about half of the WR duration for the 500 $M_\odot$.

The 200 $M_\odot$ model has an evolution  similar to the 150 $M_\odot$ model, 
while the 300 $M_\odot$ has an evolution similar to the 500 $M_\odot$.

Let us now consider how rotation changes the picture.
Figure~\ref{fig:hrdsolar} right panel shows the evolutionary tracks of the $Z=0.014$ rotating tracks in a similar way as in the left panel. The changes brought by rotation are modest. 
This is expected because of two facts: first, in this high mass range, the evolution is more impacted by mass loss than by rotation, second, stars are already well mixed by the large convective cores.
One notes however a few differences between the non-rotating and rotating models. One of the most striking is the fact that the models during their O-type phase evolve nearly vertically when rotation is accounted for. This is the effect of rotational mixing which keep the star more homogeneous than in the non-rotating cases (although, as underlined above,  already in models with no rotation, due to the importance of the convective core, stars are never very far from chemical homogeneity). As was the case in the non-rotating tracks, the O-type star phase corresponds to an upward displacement when time goes on in the HR diagram for the 150 $M_\odot$ model, while, it corresponds to a downwards displacement for the three more massive models. One notes finally that lower luminosities are reached by the rotating models at the end of their evolution (decrease by about 0.3 dex in luminosity, thus by a factor 2). This comes mainly because the rotating models enter earlier into their WR phase and thus lose more mass.

How does a change in metallicity alter the picture? 
When the metallicity decreases to Z=0.006 (see Fig.~\ref{fig:hrdlmc}), as expected, tracks are shifted to higher luminosities and effective temperatures. 
In this metallicity range, all models evolve upwards during their O-type star phase in the HR diagram. This is an effect of the lower mass loss rates.

As was already the case at Z=0.014, rotation makes the star to evolve nearly vertically in the HR diagram. One notes in this metallicity range, much more important effects of rotation than at Z=0.014, which is also expected, since at these lower metallicity, mass loss rates are smaller and rotational mixing more efficient. We note that most of the decrease in luminosity in the 500 $M_\odot$ solar mass model occurs during the WC phase in the Z=0.006 non-rotating model, while it occurs during the WNL phase in the rotating one. This illustrates the fact that rotational mixing, by creating a much larger H-rich region in the star, tends to considerably increase the duration of the WNL phase. One notes also that while the 150 $M_\odot$ model enters the WR phase only after the MS phase, the rotating model becomes a WR star before the end of the MS phase.

At the metallicity of the SMC (see Fig.~\ref{fig:hrdsmc}), except for the 500 $M_\odot$, the tracks evolve horizontally after the end of the core H-burning phase (triangle in Fig.~\ref{fig:hrdsmc}). The much lower mass loss rates are responsible for this effect.

\subsection{Lifetimes and mass-luminosity relation}

\begin{table*}
\caption{Properties of the hydrogen burning phase: initial properties of stellar models (columns 1-3), lifetime of H-burning and O-type star phase (4-5), average MS surface velocity (6) and properties at the end of the core H-burning phase (7-15). Masses are in solar masses, velocities are in km\,s$^{-1}$,
lifetimes are in $10^6$ years and abundances are surface abundances in mass fractions. The luminosity, $L$, is in log$_{10}(L/L_\odot )$ unit and the effective temperature, $T_\mathrm{eff}$, is in log$_{10}$ [K]. Note that the effective temperature given here includes a correction for WR stars to take into account the fact that their winds are optically thick as in \citet{ROTXI}.}\label{Table:endH}
\begin{tabular}{lccccccccccccccr}
\hline
 $M_{\rm ini}$  &$Z_{\rm ini}$  &$\frac{v_{\rm ini}}{v_\mathrm{crit}}$    &$\tau_H$   &$\tau_o$  &$\left<v_{\rm MS}\right>$
 &$M_\mathrm{H.b.}^\mathrm{end}$  &$^1$H  &$^4$He &$^{12}$C   &$^{14}$N &$^{16}$O &$T_\mathrm{eff}$ &$L$ &$\Gamma_\mathrm{Edd}$ \\
\hline
 120     &0.014   &0.0     &2.671   &2.592     &0.0   &63.7     &2.04e-1  &7.82e-1  &8.58e-5  &8.15e-3  &1.06e-4  &4.405 &6.334 &0.627\\
 150     &0.014   &0.0     &2.497   &2.348     &0.0   &76.3     &1.35e-1  &8.51e-1  &9.26e-5  &8.15e-3  &9.91e-5  &4.413 &6.455 &0.657 \\
 200     &0.014   &0.0     &2.323   &2.095     &0.0   &95.2     &7.51e-2  &9.11e-1  &9.93e-5  &8.14e-3  &9.23e-5  &4.405 &6.597 &0.687 \\
 300     &0.014   &0.0     &2.154   &1.657     &0.0   &65.2     &1.24e-3  &9.85e-1  &1.31e-4  &8.11e-3  &7.93e-5  &4.267 &6.401 &0.595 \\
 500     &0.014   &0.0     &1.990   &1.421     &0.0   &56.3     &2.20e-3  &9.84e-1  &1.26e-4  &8.12e-3  &8.03e-5  &4.301 &6.318 &0.568\\
 \hline
 120     &0.014   &0.4     &3.137   &2.270     &116.71 &34.6    &1.56e-3  &9.85e-1  &1.33e-4  &8.10e-3  &8.48e-5  &4.400  &6.018 &0.463 \\
 150     &0.014   &0.4     &2.909   &2.074     &101.24 &37.1    &1.80e-3  &9.85e-1  &1.30e-4  &8.11e-3  &8.41e-5  &4.387  &6.062 &0.479 \\
 200     &0.014   &0.4     &2.649   &1.830     &89.33  &40.0    &1.41e-3  &9.85e-1  &1.33e-4  &8.10e-3  &8.30e-5  &4.372  &6.110 &0.495 \\
 300     &0.014   &0.4     &2.376   &1.561     &61.16  &43.2    &1.85e-3  &9.85e-1  &1.33e-4  &8.10e-3  &8.23e-5  &4.356  &6.157 &0.511  \\
 500     &0.014   &0.4     &2.132   &1.377     &24.55  &48.1    &1.24e-3  &9.85e-1  &1.38e-4  &8.10e-3  &8.08e-5  &4.332  &6.221 &0.531 \\
 \hline
 120    &0.006    &0.0    &2.675   &2.682         &0.0  &79.0   &4.03e-1  &5.91e-1  &3.29e-5  &3.50e-3  &4.47e-5  &4.441 &6.391 &0.672 \\
 150    &0.006    &0.0    &2.492   &2.499         &0.0  &96.1   &3.28e-1  &6.67e-1  &3.58e-5  &3.50e-3  &4.25e-5  &4.483 &6.524 &0.709 \\
 500    &0.006    &0.0    &1.904   &1.636         &0.0  &238.8  &2.56e-2  &9.69e-1  &5.12e-5  &3.48e-3  &3.18e-5  &4.032 &7.094 &0.819 \\
\hline
 120     &0.006   &0.4   &3.140   &2.479         &208.55  &64.0 &1.70e-3  &9.92e-1  &6.06e-5  &3.47e-3  &3.04e-5  &4.387 &6.395 &0.597  \\
 150     &0.006   &0.4   &2.857   &2.172         &198.19  &71.3 &9.76e-4  &9.93e-1  &6.33e-5  &3.47e-3  &2.97e-5  &4.365 &6.455 &0.615  \\
 200     &0.006   &0.4   &2.590   &1.894         &193.05  &80.7 &1.22e-3  &9.93e-1  &6.29e-5  &3.47e-3  &2.95e-5  &4.339 &6.525 &0.638 \\
 300     &0.006   &0.4   &2.318   &1.619         &173.47  &85.8 &1.32e-3  &9.93e-1  &6.30e-5  &3.47e-3  &2.93e-5  &4.327 &6.559 &0.649 \\
 500     &0.006   &0.4   &2.077   &1.419         &116.76 &101.7 &1.37e-3  &9.93e-1  &6.37e-5  &3.47e-3  &2.89e-5  &4.291 &6.650 &0.676 \\
\hline
 150     &0.002   &0.4   &2.921   &2.567      &318.92  &128.8   &1.67e-3  &9.96e-1  &2.13e-5  &1.16e-3  &8.09e-6  &4.394 &6.780  &0.720 \\
 200     &0.002   &0.4   &2.612   &2.168      &333.43  &152.2   &1.31e-3  &9.97e-1  &2.26e-5  &1.16e-3  &7.82e-6  &4.363 &6.867  &0.743 \\
 300     &0.002   &0.4   &2.315   &1.801      &347.32  &176.2   &1.10e-3  &9.97e-1  &2.32e-5  &1.16e-3  &7.68e-6  &4.279 &7.067  &0.763 \\
 \hline
\end{tabular}
\end{table*}

\begin{table*}
\caption{Properties of the helium burning phase:
initial properties of stellar models (columns 1-3), age of star at the end of He-burning (4), average He-b. surface velocity (5) and properties at the end of the core He-burning phase (6-15).{ Abundances are given for the surface, except for $^{12}$C$_c$, which represents the central C abundance. Same units as in Table \ref{Table:endH}. ${\cal L}_{\rm CO}$ [$10^{50}\,\frac{{\rm g\,cm}^2}{\rm s}$] is the angular momentum contained in the CO core (Note that at this stage the angular velocity is constant in the CO core due to convective mixing).}}\label{Table:endHe}
\centering
\begin{tabular}{ccccccccccccccc}
\hline
 $M_{\rm ini}$  &$Z_{\rm ini}$  &$\frac{v_{\rm ini}}{v_\mathrm{crit}}$  &age$_\mathrm{He-b.}^\mathrm{end}$ &$\left<v_{\rm
 He-b.}\right>$ &$M_\mathrm{He-b.}^\mathrm{end}$   &$^4$He &$^{12}$C &$^{12}$C$_c$ &$^{16}$O &$^{22}$Ne &$T_\mathrm{eff}$ &$L$ &$\Gamma_\mathrm{Edd}$ & ${\cal L}_{\rm CO}$\\
\hline
 120  &0.014 &0.0 &3.003  &0.00    &30.9   &0.242  &0.458  &0.150  &0.281  &1.081e-02  &4.819  &6.117  &0.650 & 0    \\
 150  &0.014 &0.0 &2.809  &0.00    &41.3   &0.234  &0.436  &0.126  &0.312  &1.003e-02  &4.822  &6.278  &0.706 & 0    \\
 200  &0.014 &0.0 &2.622  &0.00    &49.4   &0.207  &0.408  &0.112  &0.366  &8.811e-03  &4.807  &6.377  &0.737 & 0    \\
 300  &0.014 &0.0 &2.469  &0.00    &38.2   &0.234  &0.443  &0.133  &0.305  &1.029e-02  &4.825  &6.236  &0.691 & 0    \\
 500  &0.014 &0.0 &2.314  &0.00    &29.8   &0.261  &0.464  &0.152  &0.257  &1.110e-02  &4.811  &6.095  &0.640 & 0    \\
 \hline
 120  &0.014 &0.4 &3.513  &1.58    &18.8   &0.292  &0.492  &0.195  &0.198  &1.196e-02  &4.806  &5.814  &0.533 & 1.91 \\
 150  &0.014 &0.4 &3.291  &1.18    &20.3   &0.286  &0.488  &0.187  &0.208  &1.184e-02  &4.808  &5.863  &0.551 & 1.91 \\
 200  &0.014 &0.4 &3.020  &0.50    &22.0   &0.277  &0.484  &0.180  &0.221  &1.172e-02  &4.812  &5.912  &0.570 & 1.37 \\
 300  &0.014 &0.4 &2.733  &0.13    &24.0   &0.270  &0.479  &0.172  &0.233  &1.151e-02  &4.814  &5.965  &0.589 & 0.75 \\
 500  &0.014 &0.4 &2.502  &0.03    &25.9   &0.269  &0.473  &0.164  &0.239  &1.140e-02  &4.811  &6.010  &0.606 & 0.28 \\
 \hline
 120  &0.006 &0.0 &2.993  &0.00    &54.2   &0.229  &0.391  &0.098  &0.372  &3.701e-03  &4.860  &6.424  &0.753 & 0    \\
 150  &0.006 &0.0 &2.845  &0.00    &59.7   &0.241  &0.370  &0.086  &0.380  &3.597e-03  &4.844  &6.474  &0.767 & 0    \\
 500  &0.006 &0.0 &2.182  &0.00    &94.7   &0.251  &0.392  &0.078  &0.349  &3.318e-03  &4.833  &6.711  &0.834 & 0    \\
 \hline
 120  &0.006 &0.4 &3.472  &6.84    &39.3   &0.294  &0.457  &0.132  &0.241  &4.709e-03  &4.387  &6.395  &0.692 & 16.2 \\
 150  &0.006 &0.4 &3.164  &3.67    &45.7   &0.310  &0.451  &0.122  &0.231  &4.701e-03  &4.824  &6.329  &0.767 & 14.7 \\
 200  &0.006 &0.4 &2.904  &1.33    &51.1   &0.303  &0.444  &0.114  &0.245  &4.547e-03  &4.825  &6.390  &0.738 & 9.98 \\
 300  &0.006 &0.4 &2.625  &0.35    &54.1   &0.291  &0.439  &0.110  &0.262  &4.433e-03  &4.830  &6.421  &0.748 & 5.18 \\
 500  &0.006 &0.4 &2.387  &0.13    &74.9   &0.330  &0.425  &0.090  &0.237  &4.356e-03  &4.790  &6.590  &0.798 & 4.83 \\
 \hline
 150  &0.002 &0.4 &3.193  &64.94   &106.7  &0.809  &0.153  &0.074  &0.035  &1.730e-03  &4.743  &6.766  &0.841 & 412.5\\
 200  &0.002 &0.4 &2.889  &29.88   &129.3  &0.880  &0.109  &0.066  &0.009  &1.777e-03  &4.789  &6.861  &0.863 & 355.6\\
 300  &0.002 &0.4 &2.585  &5.10    &149.8  &0.938  &0.058  &0.060  &0.001  &1.798e-03  &4.833  &6.933  &0.880 & 156.8\\
\hline
\end{tabular}
\end{table*}

\begin{table*}
\caption{Key properties at the end of the calculations (last model): initial properties of stellar models (columns 1-3), burning stage corresponding to the last model (4), total lifetime (5), average surface velocity after core He-burning (6) and properties at the last model (7-14). Same units as in Table \ref{Table:endH}.}\label{Table:endmod}
\centering
\begin{tabular}{cclcccccccccccr}
\hline
 $M_{\rm ini}$  &$Z_{\rm ini}$  &$\frac{v_{\rm ini}}{v_\mathrm{crit}}$ &last model &lifetime&$\left<v_{\rm s}\right>$ &$M_\mathrm{f}$
  &$^4$He &$^{12}$C &$^{16}$O &$^{22}$Ne &$T_\mathrm{eff}$ &$L$ &$\Gamma_\mathrm{Edd}$ \\
\hline
 120  &0.014 &0.0 &end O-b.  &3.007  &0.00     &30.8    &2.376e-01  &4.568e-01  &2.872e-01  &1.075e-02  &4.791 &6.252 &0.892  \\
 150  &0.014 &0.0 &end C-b.  &2.813  &0.00     &41.2    &2.268e-01  &4.332e-01  &3.214e-01  &9.910e-03  &4.009 &6.414 &0.969  \\
 200  &0.014 &0.0 &end C-b.  &2.625  &0.00     &49.3    &1.949e-01  &4.014e-01  &3.848e-01  &8.505e-03  &4.000 &6.486 &0.955 \\
 300  &0.014 &0.0 &end O-b.  &2.473  &0.00     &38.2    &2.309e-01  &4.418e-01  &3.088e-01  &1.025e-02  &4.479 &6.380 &0.966 \\
 500  &0.014 &0.0 &end O-b.  &2.318  &0.00     &29.8    &2.562e-01  &4.629e-01  &2.626e-01  &1.106e-02  &4.844 &6.229 &0.875 \\
 \hline
 120  &0.014 &0.4 &end O-b.  &3.517  &15.18    &18.7    &2.858e-01  &4.925e-01  &2.035e-01  &1.193e-02  &4.865 &5.907 &0.664 \\
 150  &0.014 &0.4 &end O-b.  &3.295  &0.78     &20.2    &2.797e-01  &4.884e-01  &2.138e-01  &1.181e-02  &4.867 &5.954 &0.692 \\
 200  &0.014 &0.4 &end Ne-b. &3.025  &1.88     &21.9    &2.712e-01  &4.838e-01  &2.267e-01  &1.168e-02  &4.878 &6.008 &0.714 \\
 300  &0.014 &0.4 &end O-b.  &2.737  &22.81    &23.9    &2.650e-01  &4.784e-01  &2.384e-01  &1.148e-02  &4.879 &6.056 &0.731 \\
 500  &0.014 &0.4 &end Ne-b. &2.507  &0.03     &25.8    &2.643e-01  &4.726e-01  &2.449e-01  &1.136e-02  &4.879 &6.101 &0.751 \\
 \hline
 120  &0.006 &0.0 &end He-b. &2.996  &0.00     &54.2    &2.286e-01  &3.911e-01  &3.722e-01  &3.701e-03  &4.860 &6.424 &0.753 \\
 150  &0.006 &0.0 &end He-b. &2.848  &0.00     &59.7    &2.413e-01  &3.702e-01  &3.804e-01  &3.597e-03  &4.844 &6.474 &0.768 \\
 500  &0.006 &0.0 &end He-b. &2.185  &0.00     &94.7    &2.509e-01  &3.919e-01  &3.490e-01  &3.318e-03  &4.833 &6.711 &0.834 \\
 \hline
 120  &0.006 &0.4 &end O-b.  &3.476  &27.47    &39.2    &2.889e-01  &4.567e-01  &2.465e-01  &4.687e-03  &4.896 &6.327 &0.832 \\
 150  &0.006 &0.4 &end Ne-b. &3.167  &8.39     &45.6    &3.057e-01  &4.505e-01  &2.360e-01  &4.680e-03  &4.897 &6.403 &0.852 \\
 200  &0.006 &0.4 &end O-b.  &2.907  &25.36    &51.0    &2.982e-01  &4.436e-01  &2.503e-01  &4.524e-03  &4.898 &6.460 &0.869 \\
 300  &0.006 &0.4 &end O-b.  &2.629  &0.17     &54.0    &2.856e-01  &4.383e-01  &2.681e-01  &4.400e-03  &4.907 &6.490 &0.879 \\
 500  &0.006 &0.4 &end O-b.  &2.390  &0.28     &74.8    &3.220e-01  &4.249e-01  &2.452e-01  &4.307e-03  &4.871 &6.655 &0.929 \\
 \hline
 150  &0.002 &0.4 &end O-b.  &3.196  &160.57   &106.5   &7.922e-01  &1.634e-01  &4.177e-02  &1.717e-03  &4.828 &6.810 &0.932 \\
 200  &0.002 &0.4 &end O-b.  &2.889  &187.90   &129.2   &8.730e-01  &1.131e-01  &1.137e-02  &1.773e-03  &4.867 &6.905 &0.956 \\
 300  &0.002 &0.4 &end O-b.  &2.587  &10.72    &149.7   &9.362e-01  &5.969e-02  &1.487e-03  &1.797e-03  &4.934 &6.970 &0.959 \\
\hline
\end{tabular}
\end{table*}

In Tables \ref{Table:endH}--\ref{Table:endmod}, we 
provide lifetimes at the end of core hydrogen-burning, core helium 
burning and the total stellar lifetimes, respectively. The end of a burning stage is chosen when the mass fraction of the main fuel becomes 
less than $10^{-5}$. 
We see that the MS lifetime of non-rotating models at solar metallicity ranges from 2.67\,Myr to 1.99\,Myr for initial masses ranging from 120 to 500 $M_\odot$
showing the well known fact that VMS have a very weak lifetime dependence on their initial mass.

\begin{figure}
\center
\includegraphics[width=0.5\textwidth,clip=]{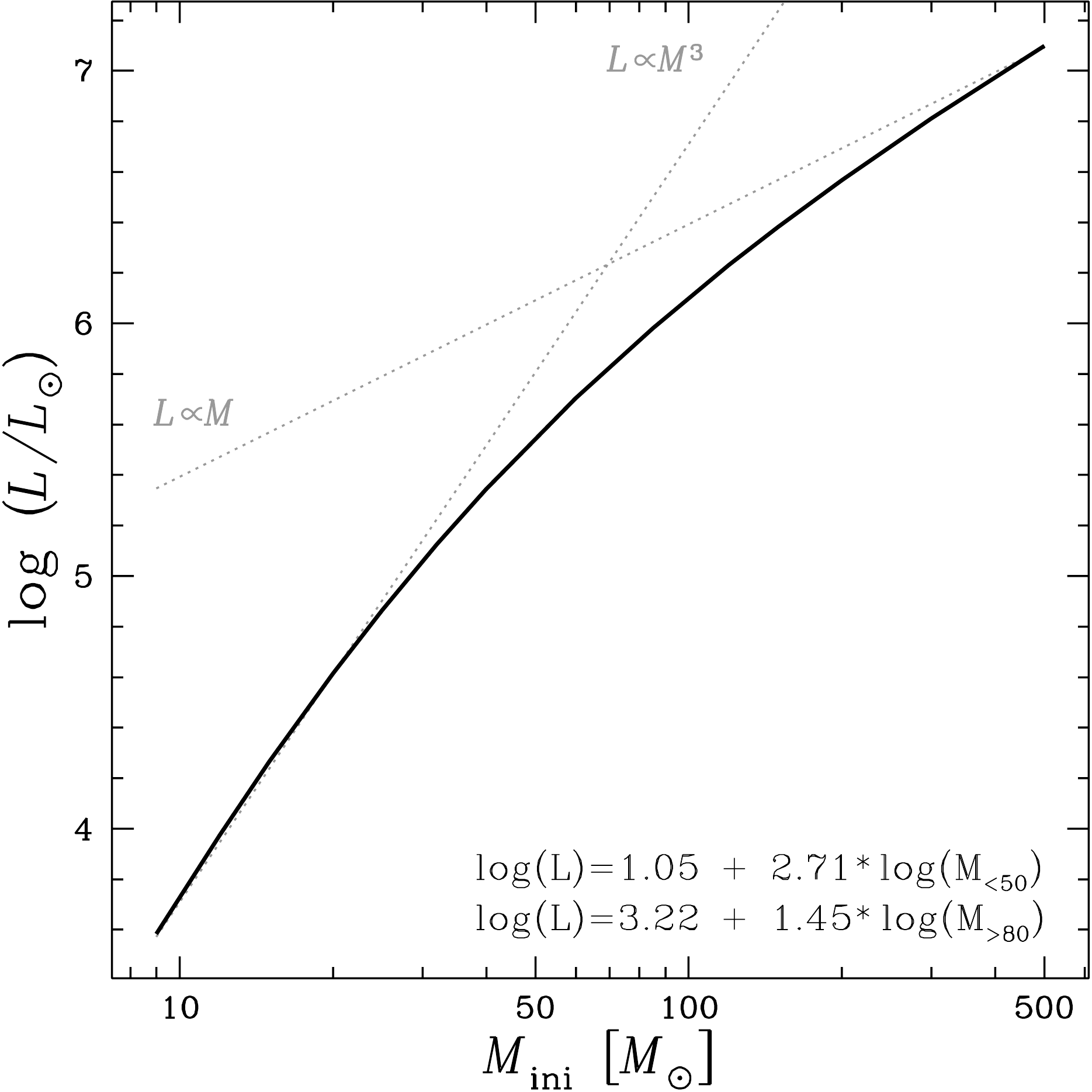}
\caption{ Mass-luminosity relation on the ZAMS for rotating models at solar metallicity. The formulae in the bottom right corner are linear fits for the mass ranges: 9--50\,$M_{\odot}$ and 80--500\,$M_{\odot}$. As can be seen in Table \ref{Table:startH}, the non-rotating models have very similar properties on the ZAMS.
}\label{fig:LvsM}
\end{figure}

The mass-luminosity relation on the ZAMS for rotating massive stars at solar 
composition is shown in Fig.~\ref{fig:LvsM}. The relation ($L \propto M^{\alpha}$) is steep for low and intermediate-mass stars ($\alpha \sim$ 3 for $10 < M/M_{\odot} < 20$) and flattens for VMS ($\alpha \sim 1.3$ for $200 < M/M_{\odot} < 500$).
This flattening is due to the increased radiation pressure relative to gas pressure in massive stars. Since the lifetime of a star is roughly $M/L$, we get that for VMS $\tau \propto M/L \propto M^{-0.3}$.

The H-burning (and total) lifetimes of VMS are lengthened by 
rotation as in lower mass stars. Differences in the H-burning lifetimes 
of rotating and non-rotating 150 $M_\odot$ models at solar metallicity 
are $\sim$ 14\%. The effects of metallicity on the lifetimes are generally very small. The small differences in total lifetimes are due to different mass loss at different metallicities.

\subsection{Mass loss by stellar winds}

\begin{figure}
\centering
\includegraphics[width=0.5\textwidth,clip=]{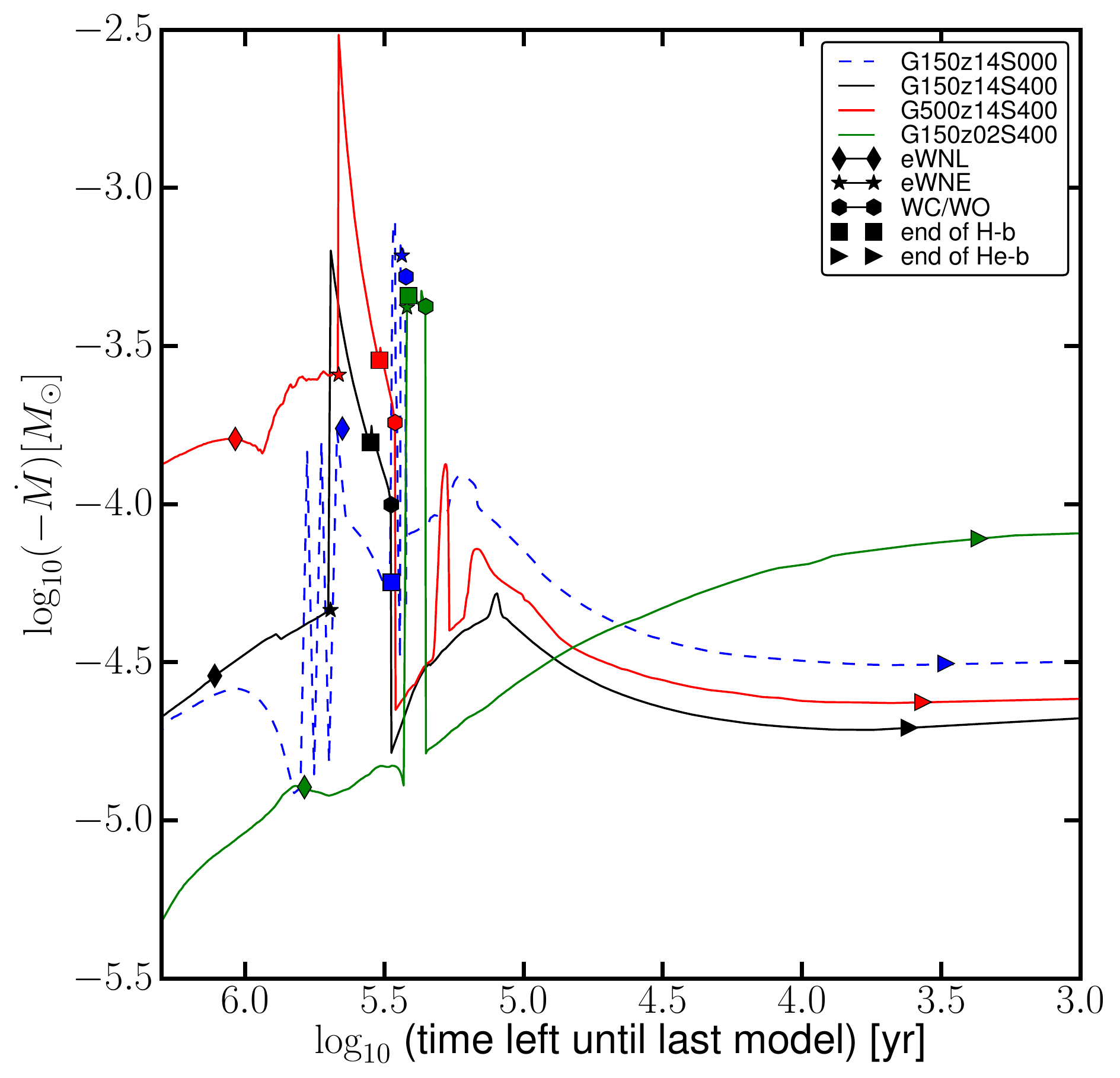}
\caption{Evolution of the mass loss rate as a function of time left until last model (log scale) for the rotating 500 $M_\odot$ model (solid-red), the rotating 150 $M_\odot$ model (solid-black), the non-rotating (dashed) 150 $M_\odot$ model at solar metallicity, and the rotating 150 $M_\odot$ model at SMC metallicity (solid-green). The diamonds indicate the start of the eWNL phase, the stars the start of the eWNE phase and hexagons the start of the WC/WO phase. The squares and triangles indicate the end of H-b. and He-b. phases, respectively.}
\label{fig:massloss_yr}
\end{figure}

\begin{table*}
\caption{Mass loss properties: Total mass of the models at various stages (columns 1-5), and average mass loss rates $\left<\dot{M}\right>$ during the O-type and eWNE phases (6,7). Masses are in solar mass units and mass loss rates are given in $M_\odot$yr$^{-1}$. }\label{mdotrates}
\centering
 \begin{tabular}{ccccc|cc}
\hline
ZAMS &end O-type/start eWNL & end eWNL/start eWNE & end eWNE/start WC & final &$\left<\dot{M}_\mathrm{Vink}\right>$ &$\left<\dot{M}_\mathrm{eWNE}\right>$ \\
\hline
&\multicolumn{3}{c}{\textbf{Z=0.014, $v/v_{crit}=0.0$}} &&&  \\
120   &69.43  &52.59  &47.62 &30.81 &2.477e-05 &3.638e-04  \\
150   &88.86  &66.87  &61.20 &41.16 &3.274e-05 &6.107e-04  \\
200   &121.06 &91.20  &83.85 &49.32 &4.618e-05 &1.150e-03  \\
300   &184.27 &130.47 &52.05 &38.15 &8.047e-05 &8.912e-04  \\
500   &298.79 &169.50 &45.14 &29.75 &1.736e-04 &9.590e-04  \\
\hline
&\multicolumn{3}{c}{\textbf{Z=0.014, $v/v_{crit}=0.4$}} &&&  \\
120   &88.28 &69.54   &27.43 &18.68 &1.675e-05 &2.057e-04   \\
150   &106.64 &80.88  &29.49 &20.22 &2.467e-05 &2.640e-04  \\
200   &137.52 &98.75  &31.84 &21.93 &3.985e-05 &3.564e-04  \\
300   &196.64 &129.10 &34.45 &23.93 &7.559e-05 &5.160e-04  \\
500   &298.42 &174.05 &38.30 &25.83 &1.594e-04 &7.901e-04  \\
\hline
&\multicolumn{3}{c}{\textbf{Z=0.006, $v/v_{crit}=0.0$}} &&&  \\
120   &74.30  &57.91  &56.91  &54.11  &2.140e-05 &3.272e-04  \\
150   &94.18  &74.20  &71.75  &59.59  &2.839e-05 &5.038e-04  \\
500   &332.68 &250.64 &197.41 &94.56  &1.304e-04 &3.334e-03   \\
\hline
&\multicolumn{3}{c}{\textbf{Z=0.006, $v/v_{crit}=0.4$}} &&&  \\
120   &100.57 &90.78  &54.43 &39.25 &9.429e-06 &3.219e-04 \\
150   &125.79 &111.84 &60.75 &45.58 &1.367e-05 &4.418e-04 \\
200   &166.81 &144.86 &66.25 &51.02 &2.180e-05 &6.257e-04 \\
300   &247.07 &207.10 &73.11 &54.04 &4.166e-05 &9.524e-04 \\
500   &397.34 &315.51 &86.10 &74.75 &9.194e-05 &1.685e-03  \\
\hline
&\multicolumn{3}{c}{\textbf{Z=0.002, $v/v_{crit}=0.4$}} &&&  \\
150  &135.06 &130.46 &113.51 &106.50 &6.661e-06 &4.485e-04 \\
200  &181.42 &174.18 &137.90 &129.21 &9.902e-06 &6.631e-04 \\
300  &273.18 &260.81 &156.14 &149.70 &1.730e-05 &1.040e-03 \\
\hline
 \end{tabular}
\end{table*}

Mass loss by stellar winds is a key factor governing the evolution of VMS. This comes from the very high luminosities reached by these objects. 
For example, the luminosity derived for R136a1 is about 10 million times that of our sun.

For such luminous objects,
winds will be very powerful at all evolutionary stages, so while
early main-sequence VMS are formally O-type stars from an evolutionary 
perspective, their spectral appearance may be closer to Of or Of/WN at 
early phases \citep{Crowther12}.

Table \ref{mdotrates} gives the total mass at the start and end of the evolution as well as at the transitions between the different WR phases in columns 1 to 5. 
The average mass loss rates during the O-type and eWNE phases (the phase during which the mass loss rates are highest)
are given in columns 6 and 7, respectively. 

The evolution of the mass loss rates for various models are shown in Fig.~\ref{fig:massloss_yr}.
Following the evolution from left to right for the 150 $M_\odot$ model at solar metallicity (solid-black), mass loss rates slowly increase at the start of the O-type phase with mass loss rates between $10^{-5}$ (absolute values for the mass loss rates, $-\dot{M}$, are quoted in this paragraph) and $10^{-4.5}$. 
If a bi-stability limit is encountered during the O-type phase, as is the case in the non-rotating 150 $M_{\odot}$ model, mass loss rates can vary significantly over a short period of time and mass loss peaks reach values higher than $10^{-4}$. The highest mass loss rate is encountered at the start of the eWNE phase (star symbols) with values in excess of $10^{-3}$ (note that the mass loss rate in the non-rotating model has a peak at the end of the H-burning phase. phase due to the star reaching temporarily cooler effective temperatures). Such high mass loss rates quickly reduce the mass and luminosity of the star and thus the mass loss rate also decreases quickly during the eWNE phase. During the WC/WO phase, mass loss rates are of the same order of magnitude as during the O-type phase.

Comparing the rotating 500 and 150 $M_\odot$ model at solar metallicity (solid black and red), we see that more massive stars start with higher mass loss rates but converge later on to similar mass loss rates since the total mass of the models converges to similar values (see Table \ref{mdotrates}). 

Comparing the SMC and solar metallicity 150 $M_\odot$ rotating models, we can clearly see the metallicity effect during the O-type star phase. During the eWNE phase, mass loss rates are similar and in the WC/WO, mass loss rates in the SMC model are actually higher since the total mass in that model remained high in contrast with solar metallicity models.

Table \ref{mdotrates} also shows the relative importance of the mass lost during the various phases and how their importance changes as a function of metallicity. Even though mass loss is the strongest during the eWNE phase, significant amount of mass is lost in all phases.

\subsection{Mass loss rates and proximity of the Eddington limit}

\begin{figure}
\centering
\includegraphics[width=0.5\textwidth,clip=]{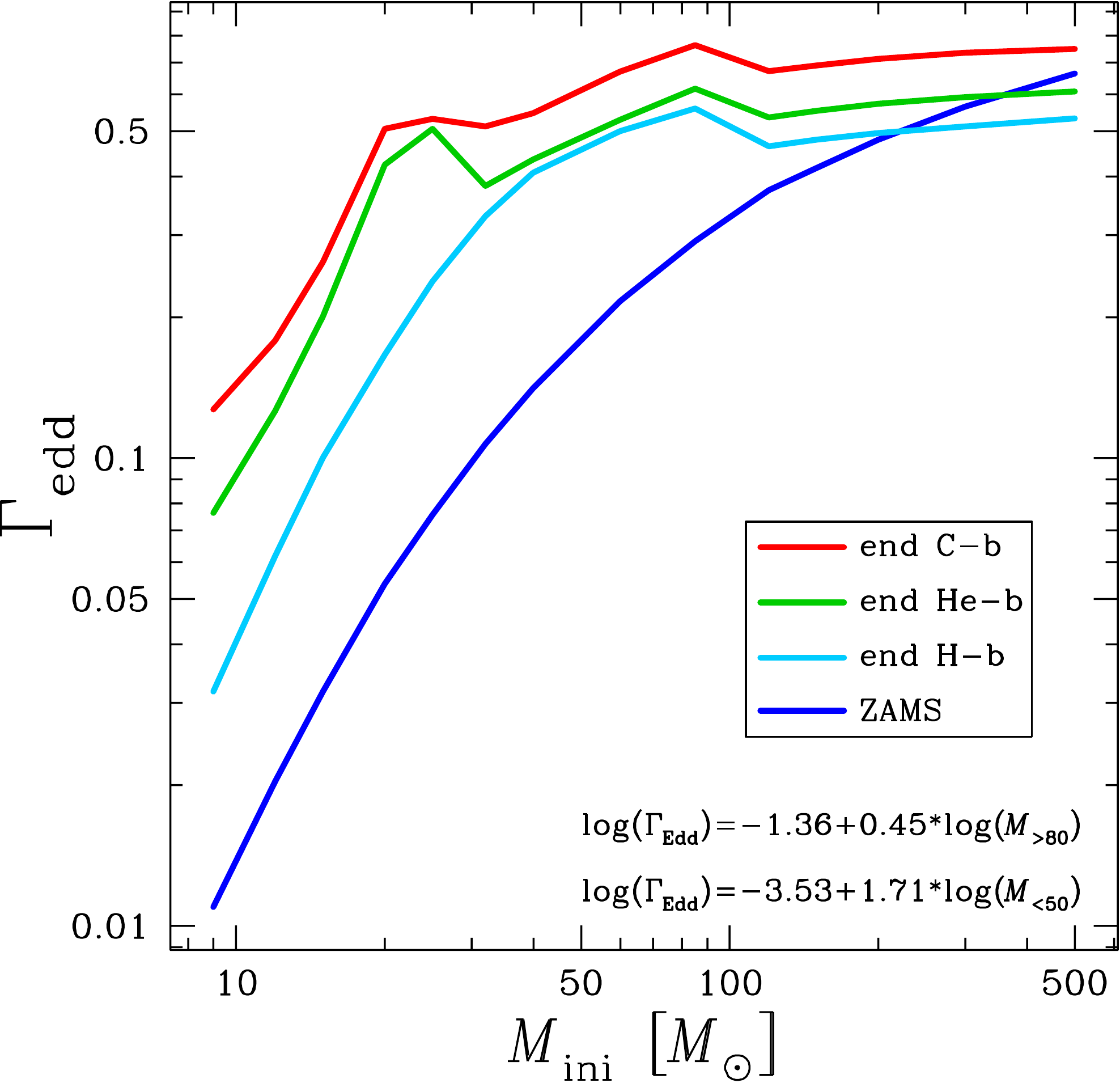}
\caption{Eddington parameter, $\Gamma_{\rm edd}$ for rotating models at solar metallicity. $\Gamma_{\rm edd}$ is plotted on the ZAMS (blue line) and the end of H-(light blue), He-(green) and C-burning (red) phases. Except for the 300 and 500 $M_\odot$ models, $\Gamma_{\rm edd}$ increases throughout the evolution. At solar metallicity, the highest value (close to 0.8) is actually reached by the 85 $M_\odot$ model at the end of its evolution. This could lead to significant mass loss shortly before the final explosion in a model that ends as a WR star and potentially explain supernova surrounded by a thick circumstellar material without the need for the star to be in the luminous variable phase.
The formulae in the bottom right corner are linear fits for the mass ranges: 9--50\,$M_{\odot}$ and 80--500\,$M_{\odot}$. As can be seen in Table \ref{Table:startH}, the non-rotating models have very similar properties on the ZAMS.
}\label{fig:edd_mass}
\end{figure}

Recently, \citet{Grafener11} suggested enhanced
mass-loss rates \citep[with respect to][]{VN01} for stars with high 
Eddington parameters ($\Gamma_{\rm Edd} \geq$ 0.7) that they attribute to the 
Wolf-Rayet stage. In the present work, we did not use such a mass loss rate prescription. In order to know whether
it would have had an impact on the present result, we discuss here the proximity of our models to the Eddington limit.

Figure \ref{fig:edd_mass} shows the Eddington parameter, $\Gamma_\text{Edd}=L/L_\text{Edd}=\kappa L / (4\pi cGM)$, as a function of the initial mass of our models at key stages 
(see also Table \ref{Table:startH}).
Since the Eddington parameter, $\Gamma_\text{Edd}$ scales with $L/M$, the curve for $\Gamma_\text{Edd}$ also flattens for VMS.
The ZAMS values for $\Gamma_\text{Edd}$ range between $0.4 - 0.6$, so well below the Eddington limit, $\Gamma_\text{Edd} = 1$, and
below the limiting value of 0.7 where enhanced
mass-loss rates are expected according to  \citet{Grafener11}. 

\begin{figure}
\centering
\includegraphics[width=0.5\textwidth,clip=]{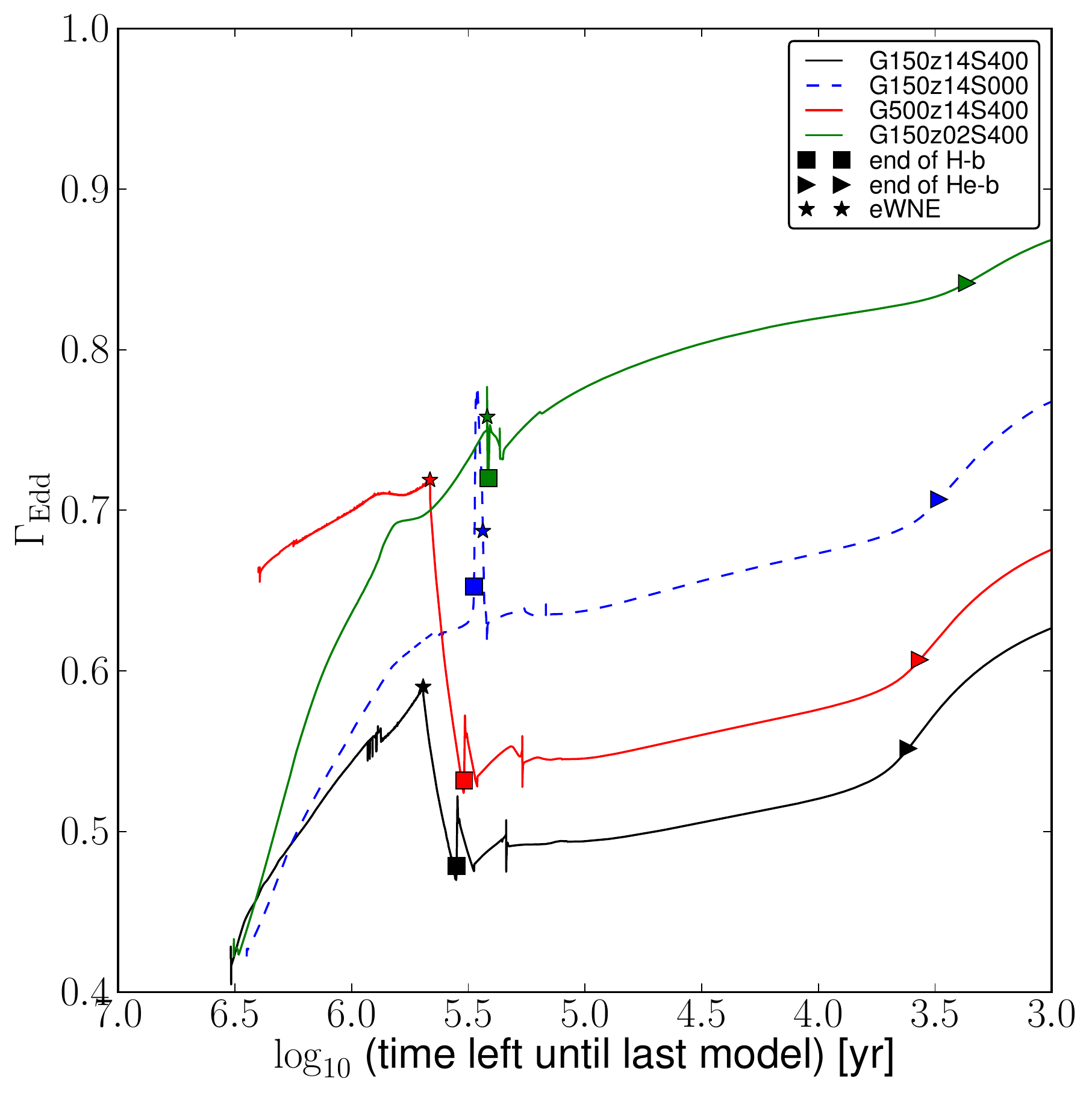}
\caption{Evolution of the Eddington parameter, $\Gamma_\text{Edd}$, as a function of time left until last model (log scale) for the rotating 500 $M_\odot$ model (solid-red), the rotating 150 $M_\odot$ model (solid-black), the non-rotating (dashed) 150 $M_\odot$ model at solar metallicity, and the rotating 150 $M_\odot$ model at SMC metallicity (solid-green). The stars indicate the start of the eWNE phase. The squares and triangles indicate the end of H-b. and He-b. phases, respectively.
}\label{fig:edd_yr}
\end{figure}

How does $\Gamma_\text{Edd}$ change during the lifetime of VMS? {Fig.~\ref{fig:edd_yr} presents the evolution of $\Gamma_\text{Edd}$ for a subset of representative models. The numerical values for each model are given at key stages in Tables \ref{Table:endH}--\ref{Table:endmod}. Since $\Gamma_\text{Edd}\propto \kappa L/M$, an increase in luminosity and a decrease in mass both lead to higher $\Gamma_\text{Edd}$. Changes in effective temperature and chemical composition affect the opacity and also lead to changes in $\Gamma_\text{Edd}$.}

{In rotating models at solar metallicity, $\Gamma_\text{Edd}$ slowly increases until the start of the eWNE phase. This is mainly due to the increase in luminosity and decrease in mass of the model. At the start of the eWNE phase, mass loss increases significantly. This leads to a strong decrease in the luminosity of the model and as a result $\Gamma_\text{Edd}$ decreases sharply.}

During the WC/WO phase, mass loss rates being of similar values as during the O-type star phase, $\Gamma_\text{Edd}$ increases again gradually.

We can see that, at solar metallicity, $\Gamma_\text{Edd}$ rarely increases beyond 0.7 even in the 500 $M_\odot$ model. 
There are nevertheless two interesting cases in which values above 0.7 are reached. The first case is during the advanced stages. At this stage, mass loss does not have much time to change the total mass of the star (it is mostly changes in effective temperature and to a minor extent in luminosity that influence the increase in $\Gamma_\text{Edd}$). This may nevertheless trigger
instabilities resulting in strong mass loss episodes.
This may have consequences for the type of SN
event that such star will produce and may be a
reason why the explosion of VMS may look like as if they
had happened in environment similar to those observed around Luminous Blue Variable. The second case is at low metallicity, as highlighted by the 150 $M_\odot$ model. Indeed, values above 0.7 are reached before the end of the MS (square symbol). We plan to determine the impact of using mass loss prescriptions such as the ones of \citet{Grafener11} and \citet{MGME12} on the fate of VMS in a forthcoming study. The non-rotating model has a different mass loss history (see Fig.~\ref{fig:massloss_yr}), which explains the slightly different evolution of $\Gamma_\text{Edd}$ near the end of the main sequence.

\subsection{Evolution of the surface velocity}\label{vsurf}

\begin{figure*}
\centering
\includegraphics[width=0.5\textwidth,clip=]{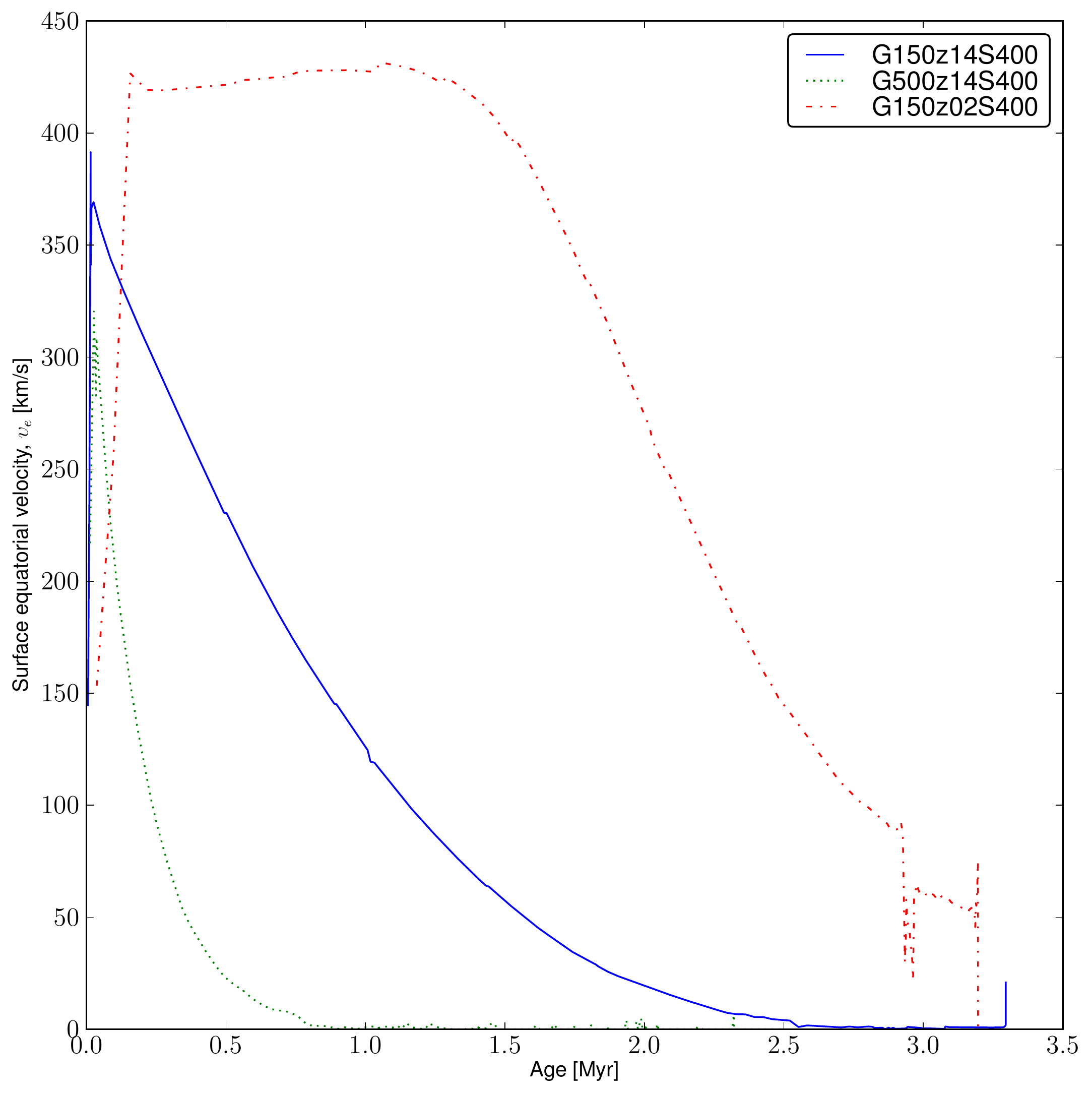}\includegraphics[width=0.5\textwidth,clip=]{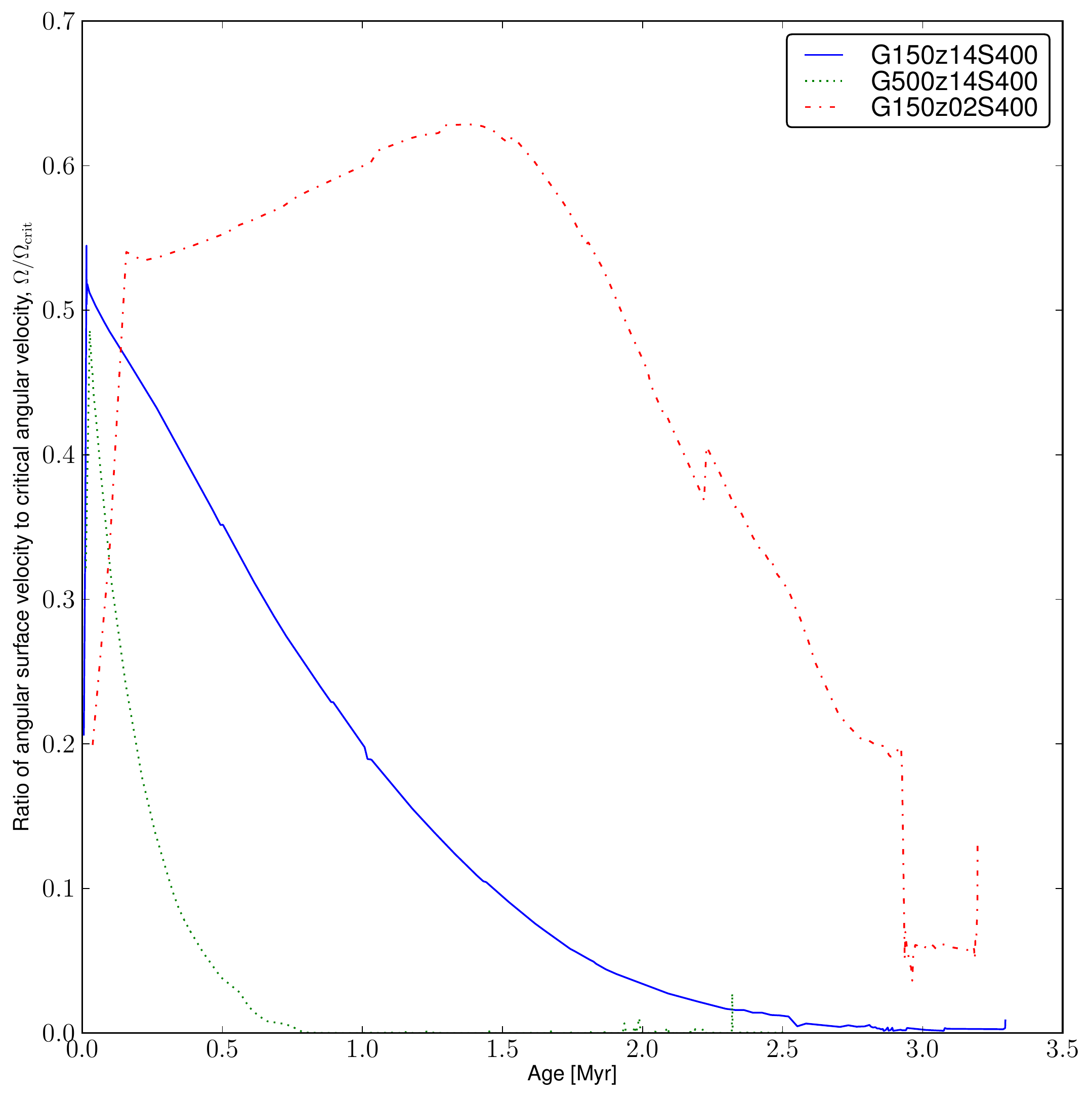} \\
\caption{Evolution of surface equatorial velocity ({\it left}) and ratio of the surface angular velocity to the critical angular velocity ({\it right}) for the rotating solar metallicity 150 and 500 $M_\odot$ and SMC 150 $M_\odot$ models as a function of age of the star.}\label{fig:vsurf}
\end{figure*}
The surface velocity of stars is affected by several processes. Contraction or expansion of the surface respectively increases and decreases the surface velocity due to conservation of angular momentum. Mass loss removes angular momentum and thus decreases the surface velocity. Finally internal transport of angular momentum generally increases the surface velocity. As shown in Fig.\,\ref{fig:vsurf} (left panel), at solar metallicity, the surface velocity rapidly decreases during the main sequence due to the strong mass loss over the entire mass range of VMS. At SMC metallicity, mass loss is weaker and internal transport of angular momentum initially dominate over mass loss and the surface velocity increases during the first half of the MS phase. During this time, the ratio of surface velocity to critical velocity also increases up to values close to 0.7 \citep[note that our models include the effect of the luminosity of the star when determining the critical rotation as described in ][]{ROTVI}. However, at SMC metallicity, in contrast with very low and zero metallicity stars \citep{H07,ES08,YDL12,CE12}, 
mass loss eventually starts to dominate and the surface velocity and its ratio to critical rotation both decrease for the rest of the evolution. SMC stars thus never reach critical rotation. The angular momentum content in the core of VMS stars are discussed in Sect. \ref{grb}

\section{WR stars from VMS}
\begin{figure}
\centering
\includegraphics[width=0.5\textwidth,clip=]{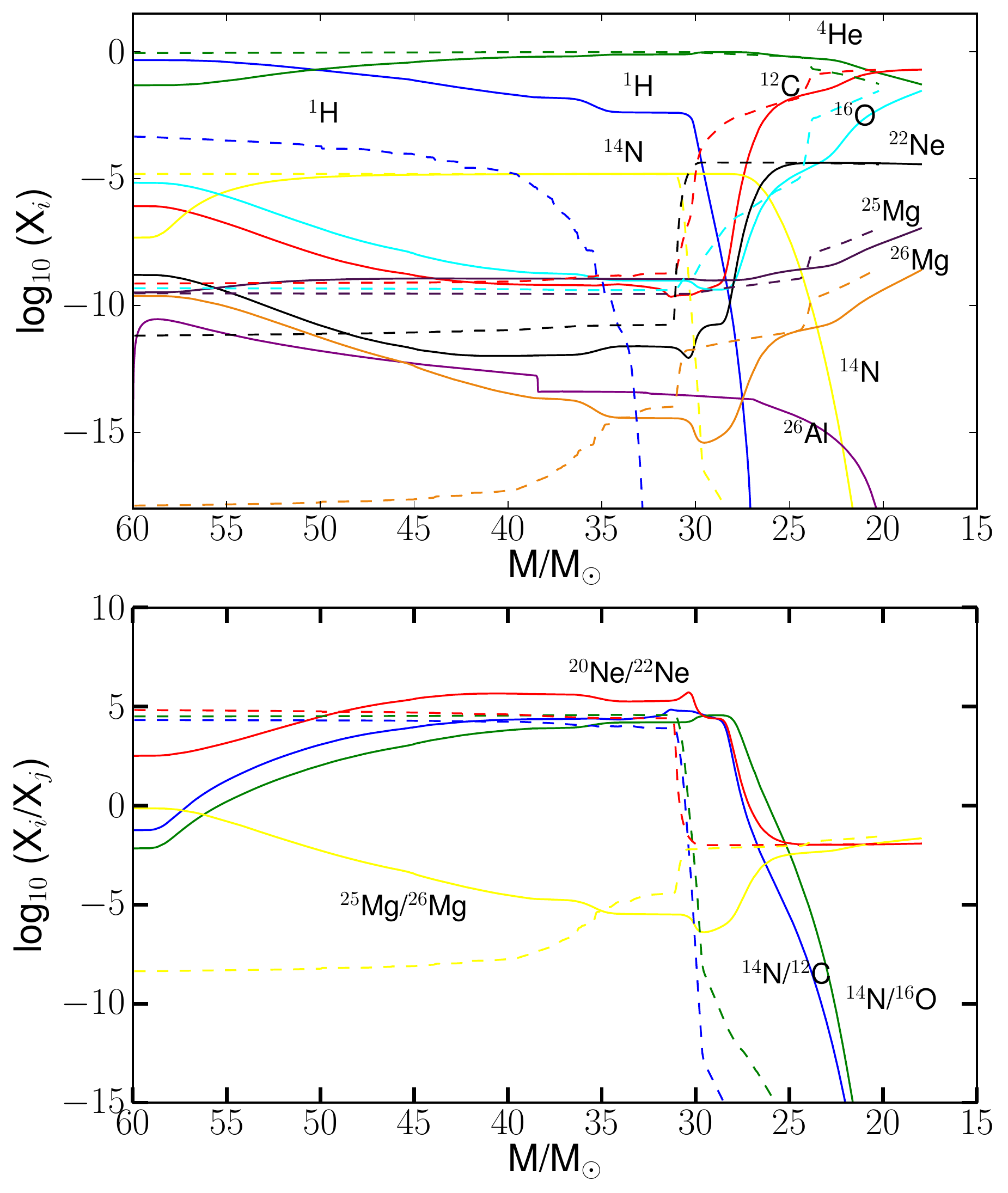} \\
\caption{
Evolution of surface abundances of the solar metallicity rotating 150 $M_\odot$ (solid) and 60 $M_\odot$ (dashed) rotating solar $Z$ models as a function of total mass (evolution goes from left to right since mass loss peels off the star and reduces the total mass). The top panel shows individual abundances while the bottom panel shows abundance ratios.]
}\label{fig:surface_solar}
\end{figure}

In Fig. \ref{fig:surface_solar}, 
we present the evolution of the surface abundances as a function of the total mass for the solar metallicity rotating models of 150 and 60 $M_\odot$.
This figure shows how the combined effects of mass loss and internal mixing change their surface composition.
Qualitatively there are no big differences between the 60 and 150 $M_\odot$ models.
Since the 150 $M_\odot$ has larger cores, the transition to the various WR stages occurs at larger total masses compared to the 60 $M_\odot$ model. It thus confirm the general idea that a more massive (thus more luminous) WR star originates from a more massive O-type star. Figure \ref{fig:surface_solar} shows that all abundances and abundance ratios are very similar for a given WR phase. it is therefore not possible to distinguish a WR originating from a VMS from its surface chemical composition (however see below).

\begin{table*}
\caption{Lifetimes of the various phases in units of years.}\label{wrlifetime}
 \begin{tabular}{ccccccccccccr}
\hline
$M_{\rm ini}$  &$Z_{\rm ini}$  &$\frac{v_{\rm ini}}{v_\mathrm{crit}}$        &O-star     &WR            &WNL        &WNE        &WN/WC      &WC (WO)            \\
\hline
120    &0.014	 &0      &2.151e06    &3.959e05    &1.150e05   &9.390e03   &2.675e02   &2.715e05          \\
150    &0.014	 &0      &2.041e06    &4.473e05    &1.777e05   &5.654e03   &7.120e02   &2.639e05          \\
200    &0.014	 &0      &1.968e06    &5.148e05    &2.503e05   &1.773e03   &4.576e02   &2.626e05       \\
300    &0.014	 &0      &1.671e06    &8.014e05    &5.051e05   &9.217e03   &2.735e03   &2.870e05          \\
500    &0.014	 &0      &1.286e06    &8.848e05    &5.804e05   &1.079e04   &3.279e03   &2.935e05       \\
\hline
120	&0.014   &0.4    &2.289e06    &1.227e06    &8.790e05   &4.118e04   &4.008e03   &3.076e05          \\
150	&0.014   &0.4    &2.105e06    &1.189e06    &8.567e05   &2.579e04   &3.649e03   &3.068e05          \\
200	&0.014   &0.4    &1.860e06    &1.164e06    &8.375e05   &2.242e04   &3.153e03   &3.042e05          \\
300	&0.014   &0.4    &1.585e06    &1.152e06    &8.315e05   &1.897e04   &2.897e03   &3.015e05          \\
500	&0.014   &0.4    &1.422e06    &1.083e06    &7.663e05   &1.830e04   &2.899e03   &2.990e05          \\
\hline
120     &0.006	 &0      &2.222e06    &2.964e05    &2.043e05   &1.302e02   &6.025e02   &9.202e04          \\
150     &0.006	 &0      &2.028e06    &3.320e05    &1.579e05   &1.211e03   &2.921e02   &1.728e05          \\
500     &0.006	 &0      &1.388e06    &5.362e05    &2.690e05   &5.211e03   &1.350e03   &2.620e05          \\
\hline
120    &0.006    &0.4    &2.513e06    &9.624e05    &6.776e05   &1.601e04   &3.386e03   &2.687e05          \\
150    &0.006    &0.4    &2.188e06    &9.789e05    &6.912e05   &2.172e04   &2.336e03   &2.660e05          \\
200    &0.006    &0.4    &1.922e06    &9.848e05    &7.073e05   &1.347e04   &2.757e03   &2.640e05          \\
300    &0.006    &0.4    &1.644e06    &9.838e05    &7.033e05   &1.600e04   &9.744e02   &2.644e05          \\
500    &0.006    &0.4    &1.461e06    &9.283e05    &6.647e05   &9.312e03   &6.853e02   &2.542e05          \\
\hline
150    &0.002    &0.4    &2.583e06    &6.119e05    &3.691e05   &8.459e03   &4.874e03   &2.343e05          \\
200    &0.002    &0.4    &2.196e06    &6.926e05    &4.524e05   &1.019e04   &2.709e03   &2.300e05          \\
300    &0.002    &0.4    &1.827e06    &7.602e05    &5.186e05   &1.317e04   &1.289e03   &2.283e05          \\
\hline
 \end{tabular}
\end{table*}

\begin{figure*}
\center
\includegraphics[width=0.5\textwidth,clip=]{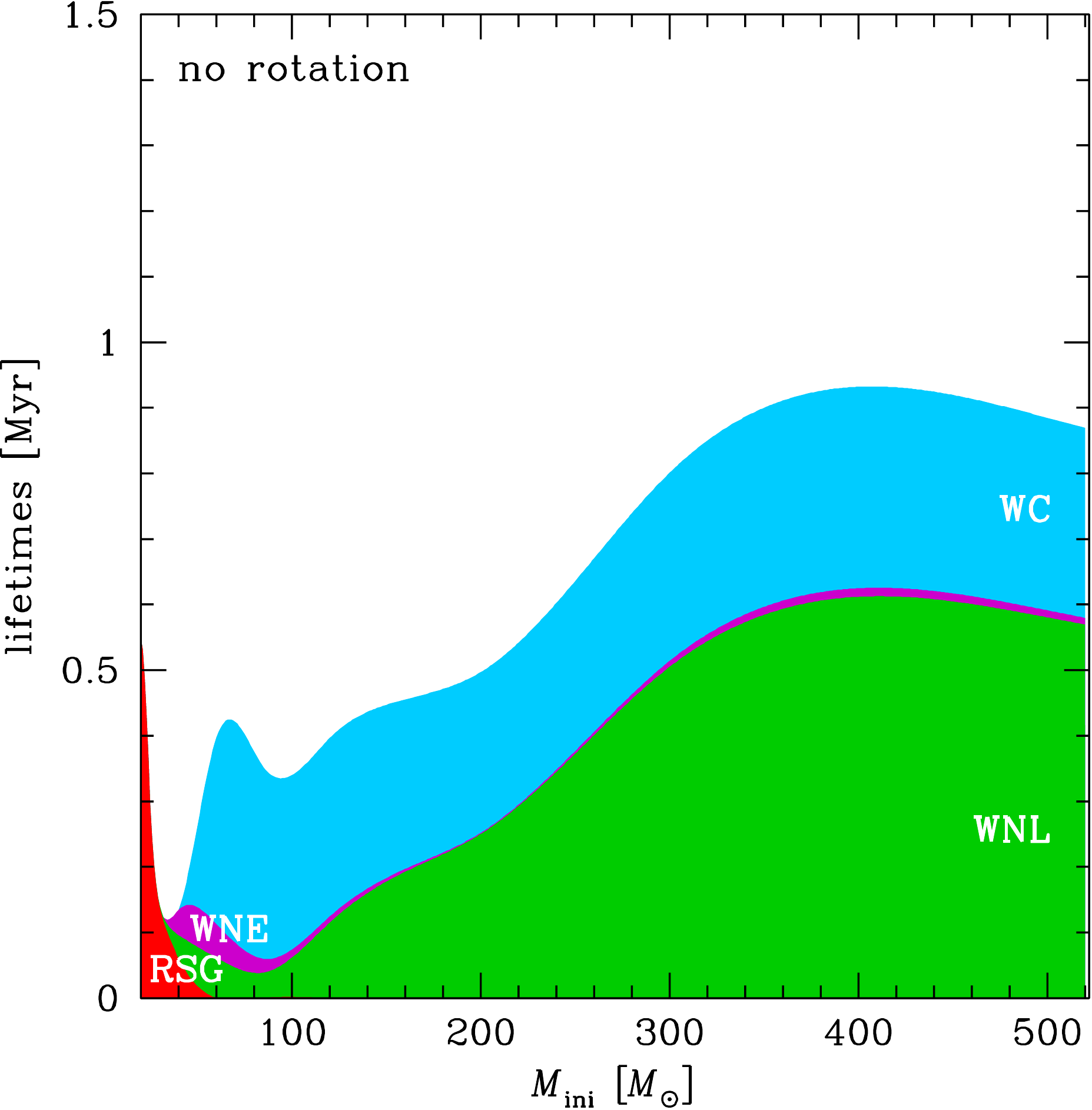}\includegraphics[width=0.5\textwidth,clip=]{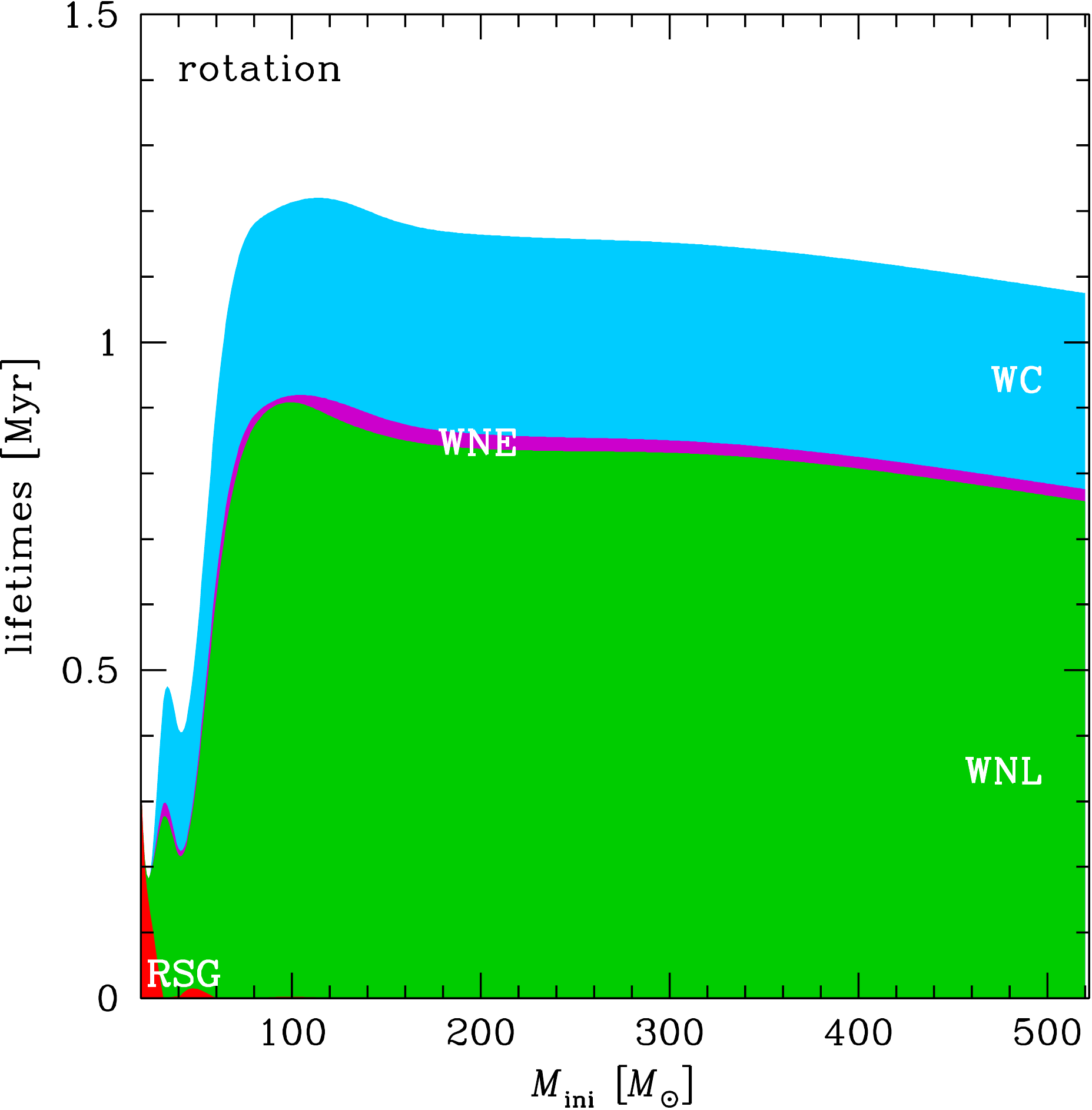}
\caption{Lifetimes of the RSG phase and of the different WR phases for the solar metallicity non-rotating ({\it left}) and rotating ({\it right}) models. Lifetimes are piled up. For example, the lifetime of the WNE phase extent corresponds to the height of the purple area.}
\label{fig:wrlife}
\end{figure*}

We present in Table \ref{wrlifetime} the lifetimes of the different WR phases
through which all our VMS models evolve.
At solar metallicity, the WR phase of non-rotating stellar models for masses between 150 and 500 $M_\odot$
covers between 16 and 38\% of the total stellar lifetime. This is a significantly larger proportion than for masses
between 20 and 120 $M_\odot$, where the WR phase covers only 0-13\% percents of the total stellar lifetimes.
At the LMC metallicity, the proportion of the total stellar lifetime spent as a WR phase for VMS decreases
to values between 12\% (150 $M_\odot$) and 25\% (500 $M_\odot$). 

Figure~\ref{fig:wrlife} shows how these lifetimes varies as a function of mass for our
non-rotating and rotating solar metallicity models.
Looking first at the non-rotating models (Fig.~\ref{fig:wrlife}, {\it left}), we see that the very massive stars
(above 150 $M_\odot$) have WR lifetimes between 0.4 and nearly 1 My.
The longest WR phase is the WNL phase since these stars spend a large fraction of H-burning in this phase. The duration
of the WC phases of VMS is not so much different from those of stars in the mass range 
between 50 and 120 $M_\odot$. 

Rotation significantly increases the WR lifetimes. Typically, the WR phase of rotating stellar models for masses between 150 and 500 $M_\odot$
covers between 36 and 43\% of the total stellar lifetime.
The increase is more important
for the lower mass range plotted in the figures. This reflects the fact that
for lower initial mass stars, mass loss rates are weaker and thus the mixing
induced by rotation has a greater impact. We see that this increase is mostly due
to longer durations for the WNL phase, the WC phase duration remaining more or less
constant for the whole mass range between 50 and 500 $M_\odot$ as was the case
for the non rotating models. Rotation has qualitatively similar effects at the LMC metallicities.

Would the account of the VMS stars in the computation of the number ratios of WR to O-type stars and on the WN/WC
ratios have a significant effect? The inclusion of VMS is marginal at solar metallicity, since the durations are only affected
by a factor 2. Convoluted with the weighting of the initial mass function (IMF), WR stars originating from VMS only represent $\sim 10\%$ of the whole
population of WR stars \citep[using a][IMF]{S55} originating from single stars. However, the situation is different at SMC metallicity. Due to the weakness of the stellar winds, single stellar models below $120\, M_
\odot$ at this $Z$ do not produce any WC or WO stars (Georgy et al. in prep.). In that case, we expect that the few WC/WO
stars observed at low metallicity come from VMS, or from the binary channel \citep{EIT08}.
In starburst regions, the detection of WR stars at very young ages would also be an indication that they come from VMS, as
these stars enter the WR phase before their less massive counterparts, and well before WRs coming from the binary
channel.
\begin{figure}
\center
\includegraphics[width=0.5\textwidth,clip=]{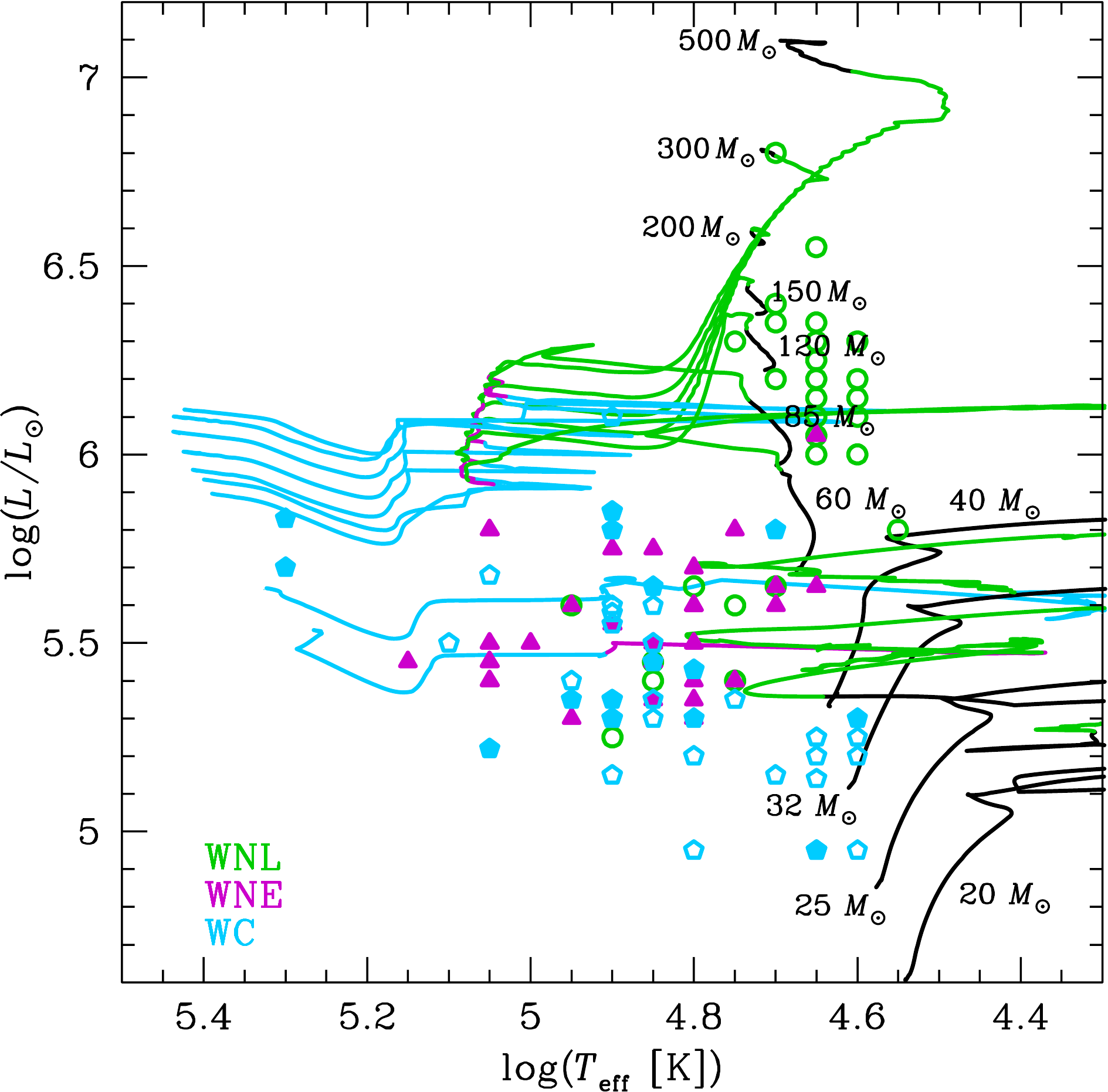}
\caption{The positions of WR stars observed by \citet{HGL06}
and \citet{SHT12} are indicated with the rotating evolutionary tracks taken from \citet{SE12} for masses up to 120 $M_\odot$ and from the present work above.
}
\label{fig:wrobs}
\end{figure}

We see in Fig.~\ref{fig:wrobs} that the present VMS models well fit the most luminous WNL stars. On the other hand, they
predict very luminous WC stars. Of course the fact that no such luminous WC stars has ever been observed can simply come from the fact that such stars are very rare and the lifetime in the WC phase is moreover relatively short.

\subsection{The final chemical structure}

\begin{figure}
\center
\includegraphics[width=0.5\textwidth,clip=]{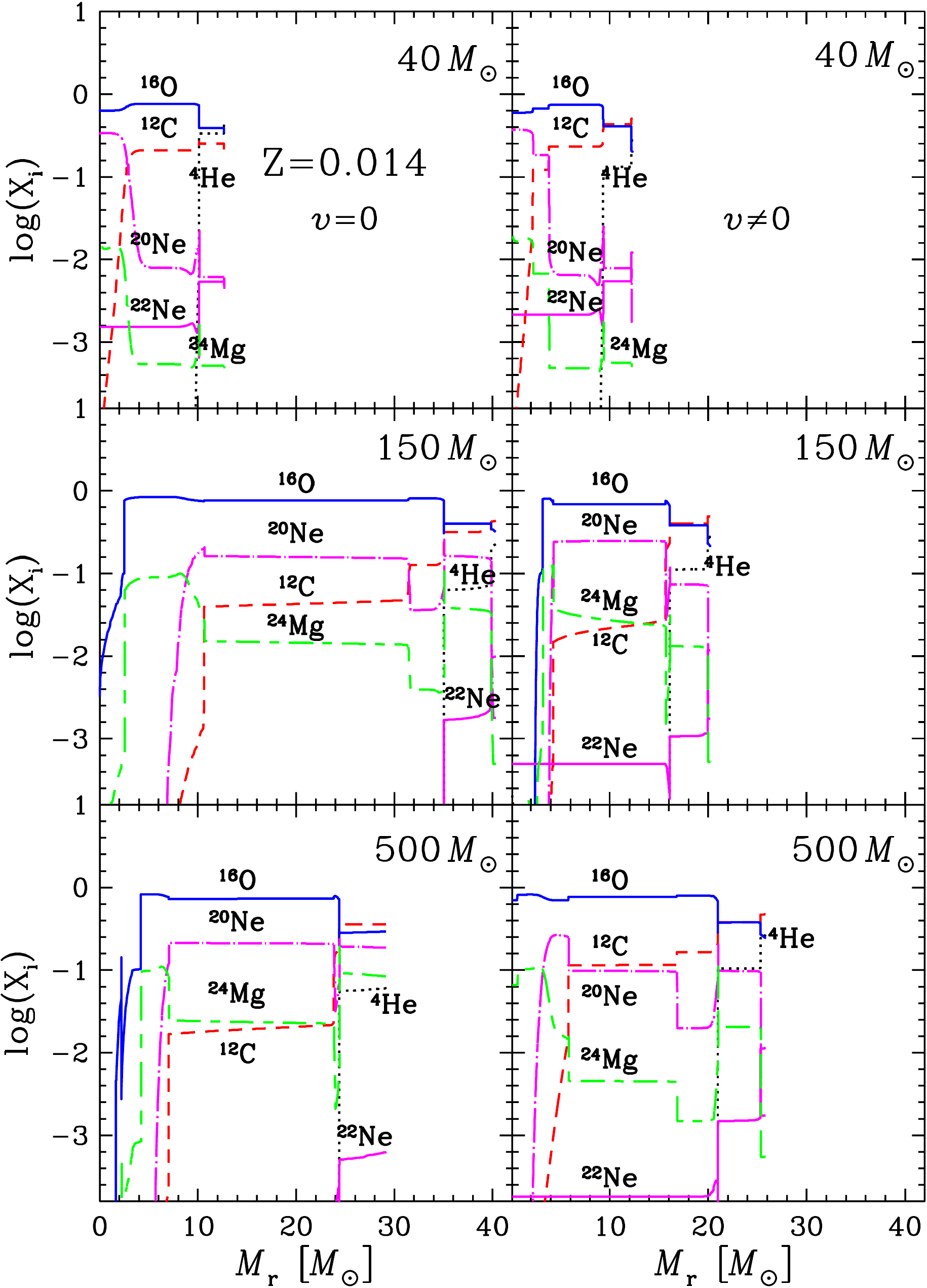}
\caption{Chemical structure of 40, 150 and 500 $M_\odot$ non-rotating ({\it left}) and rotating ({\it right}) models at Z=0.014 at the end of the calculations. Note that the rotating 500 M model is shown at an earlier evolutionary stage than the corresponding non-rotating model.
}
\label{fig:finalc}
\end{figure}

Figure \ref{fig:finalc} shows the chemical structure at the last time steps calculated, which is the end of the carbon burning phase in the case of the 40 $M_\odot$, and the end of the core oxygen-burning phase in the case of the 150 and 500 $M_\odot$ models.
A few interesting points come out from considering this figure. First, in all cases, some helium is still present in the outer layers. Depending on how the final stellar explosion occur, this helium may or may not be apparent in the spectrum, as discussed in Sect.\,\ref{fate}. 
Second, just below the He-burning shell, products of the core He-burning, not affected by further carbon burning are apparent. This zone extends between about 4 and 10 $M_\odot$ in the 40 $M_\odot$ model, between about 32 and 35 $M_\odot$ in the 150 $M_\odot$ model and in a tiny region centered around 24 $M_\odot$ in the 500 $M_\odot$ model. 
We therefore see that this zone decreases in importance when the initial mass increases. Interestingly, the chemical composition in this zone present striking differences if we compare for instance the 40 $M_\odot$ and the 500 $M_\odot$ model. We can see that the abundance of $^{20}$Ne is much higher in the more massive model. This comes from the fact that in more massive stars, due to higher central temperatures during the core He-burning phase the reaction $^{16}$O($\alpha$, $\gamma$)$^{20}$Ne is more active, building thus more $^{20}$Ne. Note that $^{24}$Mg is also more abundant, which is natural since the reaction $^{20}$Ne($\alpha$, $\gamma$)$^{24}$Mg reaction will also be somewhat active in VMS for the same reasons. While in the case of the 150 $M_\odot$, due to the mass loss history, the $^{20}$Ne and $^{24}$Mg-rich layers are not uncovered, they are uncovered in the 500 $M_\odot$ model. This implies that strong overabundances of these two isotopes at the surface of WC stars can be taken as a signature for an initially very massive stars as the progenitor of that WC star. It means also that, contrary to what occurs at the surface of WC stars originating from lower initial mass stars, neon is no longer present mainly in the form of $^{22}$Ne (and thus be a measure of the initial CNO content since resulting from the transformation of nitrogen produced by CNO burning during the H-burning phase) but will mainly be present in the form of $^{20}$Ne.
 
Rotation does not change much this picture (see right panel of Fig.~\ref{fig:finalc}), except that, due to different mass loss histories,
the rotating models lose much more mass and end their evolution with smaller cores. This is particularly striking for the 150 $M_\odot$ model.
Qualitatively the situation is not much different at lower metallicities.

\section{Fate of Very Massive Stars}\label{fate}

The best way to predict the fate of our models would be to simulate their evolution and final explosion, which we have done for a subset of the models with the KEPLER code (as described below) but {whether or not a star produces a PCSN} can be reasonably {estimated} from the mass of its carbon-oxygen (CO) core as demonstrated by the similar fate for stars with the same CO core found in various studies of VMS in the early Universe \citep{BAC84,HEGER02,CE12,DWL13}, even if their prior evolution is different. Our models also confirm the idea that the CO core mass is a good indicator for the advanced evolution and thus the fate of the models. Indeed, as discussed below, stars with a wide range of initial masses at solar metallicity end with a very similar total mass at the end of He-burning (thus a very similar CO core mass) and they have extremely similar evolution during the advanced stages. In this section, we will thus use the CO core mass {to estimate} the fate of the models. {It is important to stress that for lower-mass massive stars ($\lesssim 50\,M_\odot$), the CO core mass alone is not sufficient to predict the fate of the star and other factors like compactness, rotation and the central carbon abundance at the end of helium burning also play a role \citep[see e.\,g.][]{CL13}.} We will also discuss the supernova types that these VMS may produce.

\begin{table*}
\caption{Initial masses, mass content of helium in the envelope, mass of carbon-oxygen core, final mass in solar masses and fate of the models.}\label{table:fate}
\centering
\begin{tabular}{ccccc|ccccc}
\hline
&\multicolumn{4}{c|}{non-rotating}  &\multicolumn{4}{c}{rotating}\\
$M_{ini}$  &$M_{\textrm{He}}^{\textrm{env}}$  &$M_{\textrm{co}}$ &$M_\textrm{final}$ &Fate   &$M_{\textrm{He}}^{\textrm{env}}$  &$M_{\textrm{co}}$ &$M_\textrm{final}$ &Fate\\
\hline
\multicolumn{9}{c}{\textbf{Z=0.014}}  \\
120  &0.4874  &25.478 &30.8 &CCSN/BH   &0.5147  &18.414  &18.7 &CCSN/BH\\
150  &0.6142  &35.047 &41.2 &CCSN/BH   &0.5053  &19.942  &20.2 &CCSN/BH\\
200  &0.7765  &42.781 &49.3 &CCSN/BH   &0.5101  &21.601  &21.9 &CCSN/BH\\
300  &0.3467  &32.204 &38.2 &CCSN/BH   &0.4974  &19.468  &23.9 &CCSN/BH\\
500  &0.3119  &24.380 &29.8 &CCSN/BH   &0.5675  &20.993  &25.8 &CCSN/BH\\
\\
\multicolumn{9}{c}{\textbf{Z=0.006}}  \\
120  &1.2289  &43.851 &54.2 &CCSN/BH   &0.5665  &32.669  &39.2 &CCSN/BH\\
150  &1.1041  &47.562 &59.7 &CCSN/BH   &0.7845  &38.436  &45.6 &CCSN/BH\\
200  &-       &-      &-    &CCSN/BH   &0.5055  &42.357  &51.0 &CCSN/BH\\
300  &-       &-      &-    &CCSN/BH   &0.5802  &44.959  &54.0 &CCSN/BH\\
500  &1.6428  &92.547 &94.7 &PCSN      &0.7865  &73.145  &74.8 &PCSN\\
\\
\multicolumn{9}{c}{\textbf{Z=0.002}}  \\
150  &-  &-  &-     &-      &2.3353  &93.468  &106.5 &PCSN\\
200  &-  &-  &-     &-      &3.3022  &124.329  &129.2 &PCSN\\
300  &-  &-  &-     &-      &5.5018  &134.869  &149.7 &BH\\

\hline
\end{tabular}
\end{table*}

\subsection{Advanced phases, final masses and masses of carbon-oxygen cores}\label{tcrhoc}

\begin{figure*}
\centering
\begin{tabular}{cc}
\includegraphics[width=0.4\textwidth,clip=]{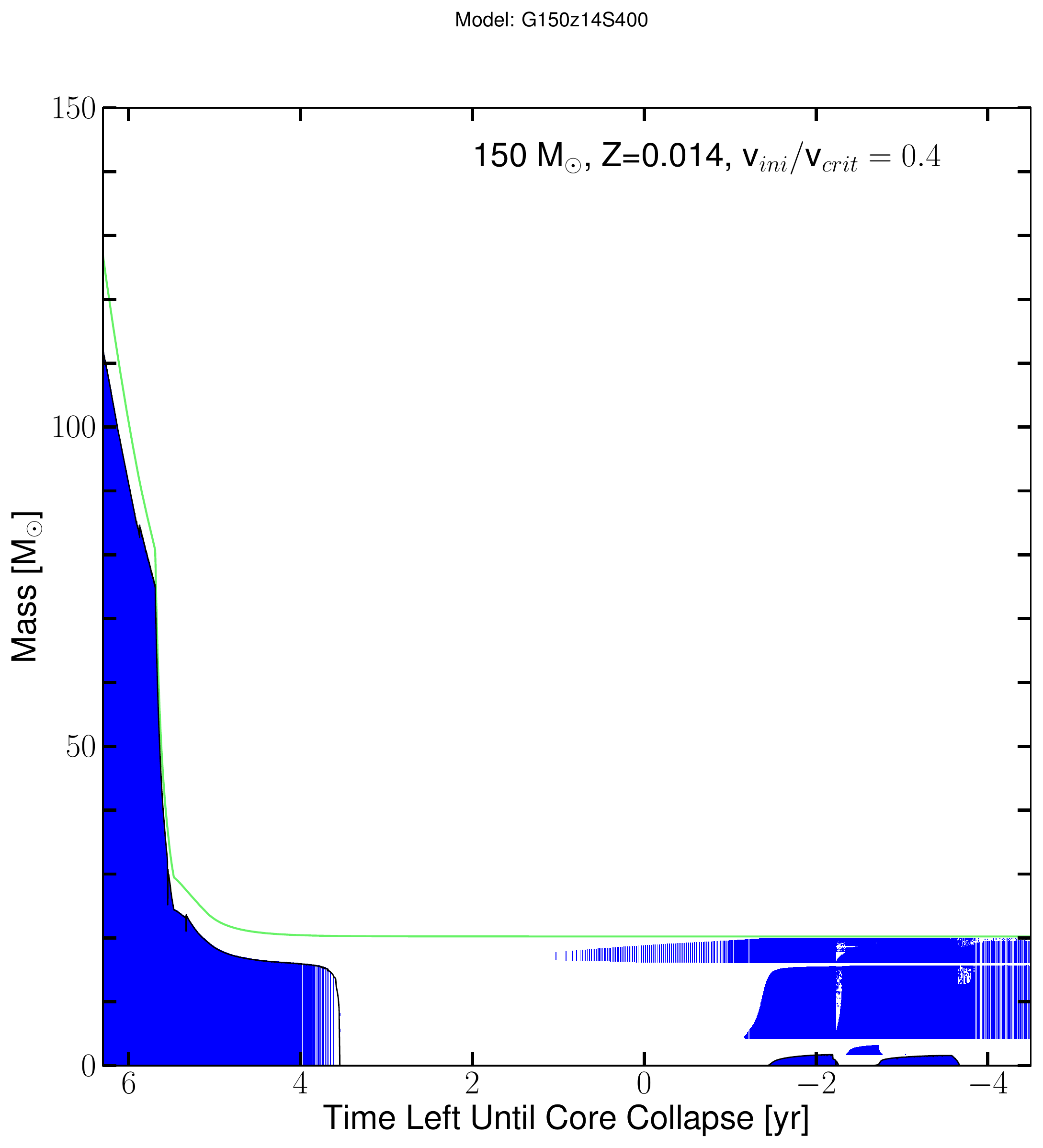} &
\includegraphics[width=0.4\textwidth,clip=]{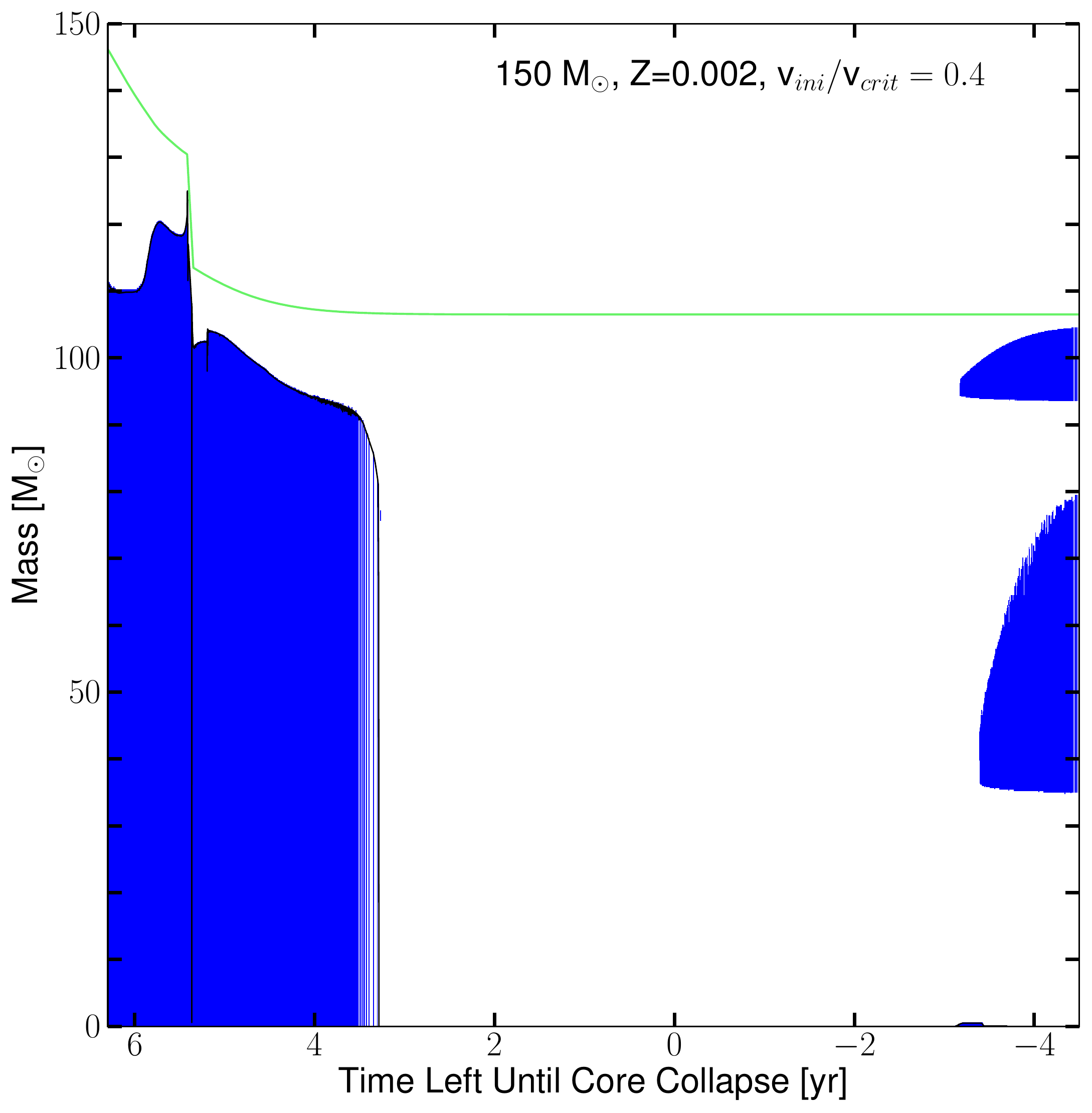} 
\end{tabular}
\caption{Structure evolution diagram for rotating 150 $M_\odot$ at solar and SMC metallicities as a function of the log of the time left until the last model. The blue zones represent the convective regions. }\label{fig:kip_logt}
\end{figure*}

In Fig. \ref{fig:kip_logt}, the structure evolution diagrams are drawn as a function of the log of the time left until the last model calculated (as opposed to age used in Fig.\,\ref{fig:kip_age}). This choice of x axis allows one to see the evolution of the structure during the advanced stages. In the {\it left} panel, we can see that, at solar metallicity, VMS have an advanced evolution identical to lower mass stars \citep[see e.\,g. Fig. 12 in][]{psn04} with a radiative core C-burning followed by a large convective C-burning shell, radiative neon burning and convective oxygen and silicon burning stages. All the solar metallicity models will eventually undergo core collapse after going through the usual advanced burning stages. {As presented in Table \ref{Table:endHe} (column 9), the central mass fraction of $^{12}$C is very low in all VMS models and is anti-correlated with the total mass at the end of helium burning (column 6): the higher the total mass, the lower the central $^{12}$C mass fraction. This is due to the higher temperature is more massive cores leading to a more efficient $^{12}$C($\alpha, \gamma$)$^{16}$O relative to 3$\alpha$.}

\begin{figure}
\center
\begin{tabular}{ccc}
\includegraphics[width=0.5\textwidth,clip=]{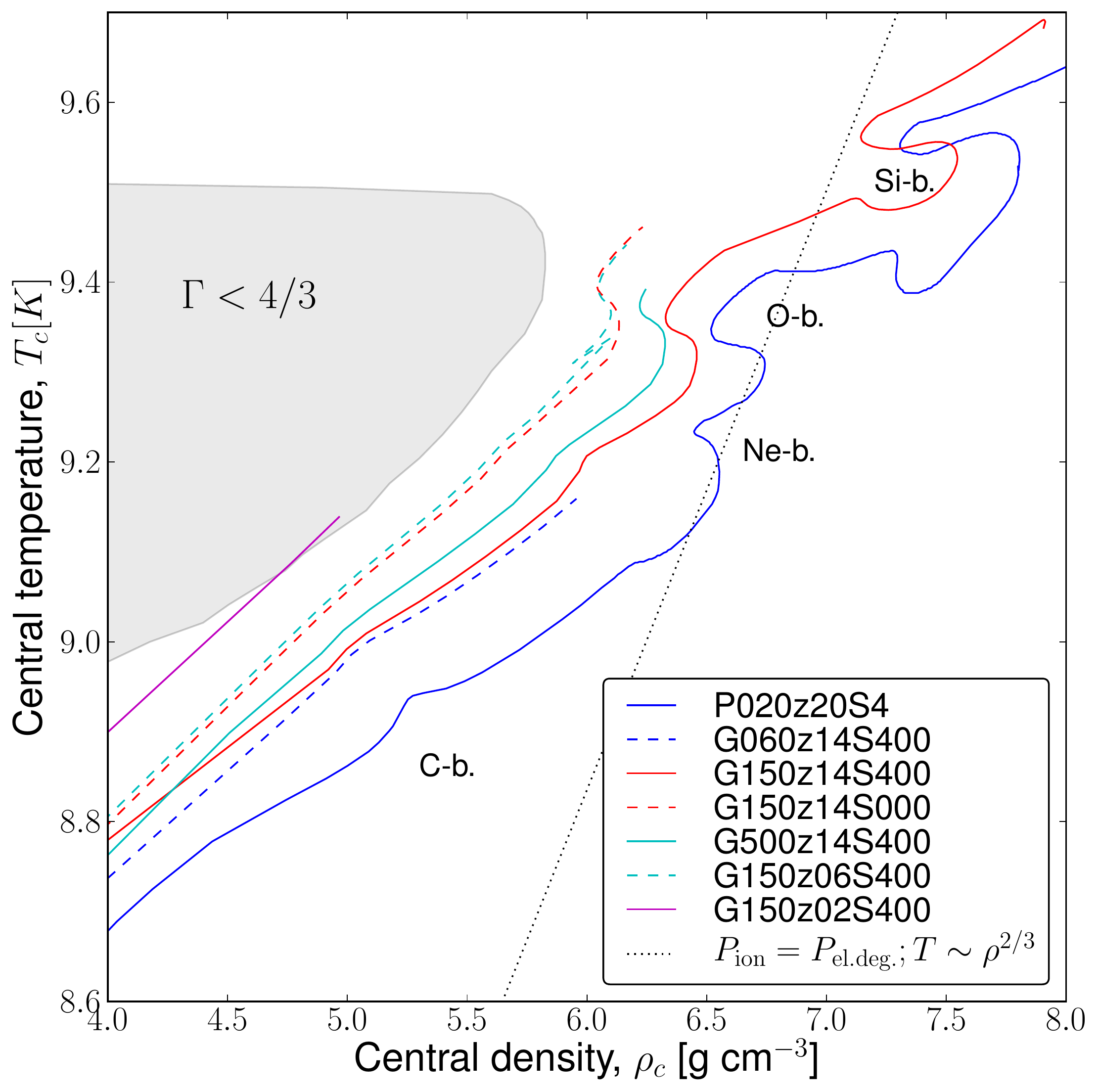} 
\end{tabular}
\caption{Evolution of the central temperature $T_c$ versus central density $\rho_c$ 
for the rotating 20 \citep[from][]{psn04}, 60 \citep[from][]{SE12}, 150 and 500 $M_\odot$ models and non-rotating 150 $M_\odot$ model at solar metallicity as well as the rotating 150 $M_\odot$ model at SMC 
metallicity. The gray shaded area is the pair-creation instability region 
($\Gamma < 4/3$, where $\Gamma$ is the adiabatic index). The additional dotted line corresponds to the limit 
between non-degenerate and degenerate electron gas.
} \label{fig:tcrhoc}
\end{figure}

The similarities between VMS and lower mass stars {at solar metallicity} during the advanced stages can also be seen in the central temperature versus central density diagram (see Fig.~\ref{fig:tcrhoc}). Even the evolution of the 500 $M_\odot$ rotating model is close to that of the 60 $M_\odot$ model. The non-rotating models lose less mass as described above and thus their evolutionary track is higher (see e.\,g. the track for the non-rotating 150 $M_\odot$ model in Fig.\,\ref{fig:tcrhoc}). Non-rotating models nevertheless stay clear of the pair-instability region ($\Gamma < 4/3$, where $\Gamma$ is the adiabatic index) in the centre.

The situation is quite different at SMC metallicity ({\it right} panel). Mass loss is weaker and thus the CO core is very large (93.5 $M_\odot$ for this 150 $M_\odot$ model). Such a large core starts the advanced stages in a similar way: radiative core C-burning followed by a large convective C-burning shell and radiative neon burning. The evolution starts to diverge from this point onwards. As can be seen in $T_c$ vs $\rho_c$ plot, the SMC 150 $M_\odot$ model enters the pair-instability region. These models will thus have a different final fate than those at solar metallicity (see below).

Figure~\ref{fig:mfinal} (see also Table \ref{mdotrates}) shows the final masses obtained in the present models as a function of the initial masses.
All models at solar $Z$, rotating or not, end with a small fraction of their initial mass due to the strong mass losses they experience.
Rotation enhances mass loss by allowing the star to enter the WR phase earlier during the MS (see {\it top} panels of Fig.~\ref{fig:kip_age}) and the final mass of non-rotating models is generally higher than that of rotating models. 
At low metallicities, due to the metallicity dependence of radiatively-driven stellar winds in both O-type stars \citep{VN01} and WR stars \citep{EV06}, final masses are larger.

\begin{figure}
\includegraphics[width=0.5\textwidth]{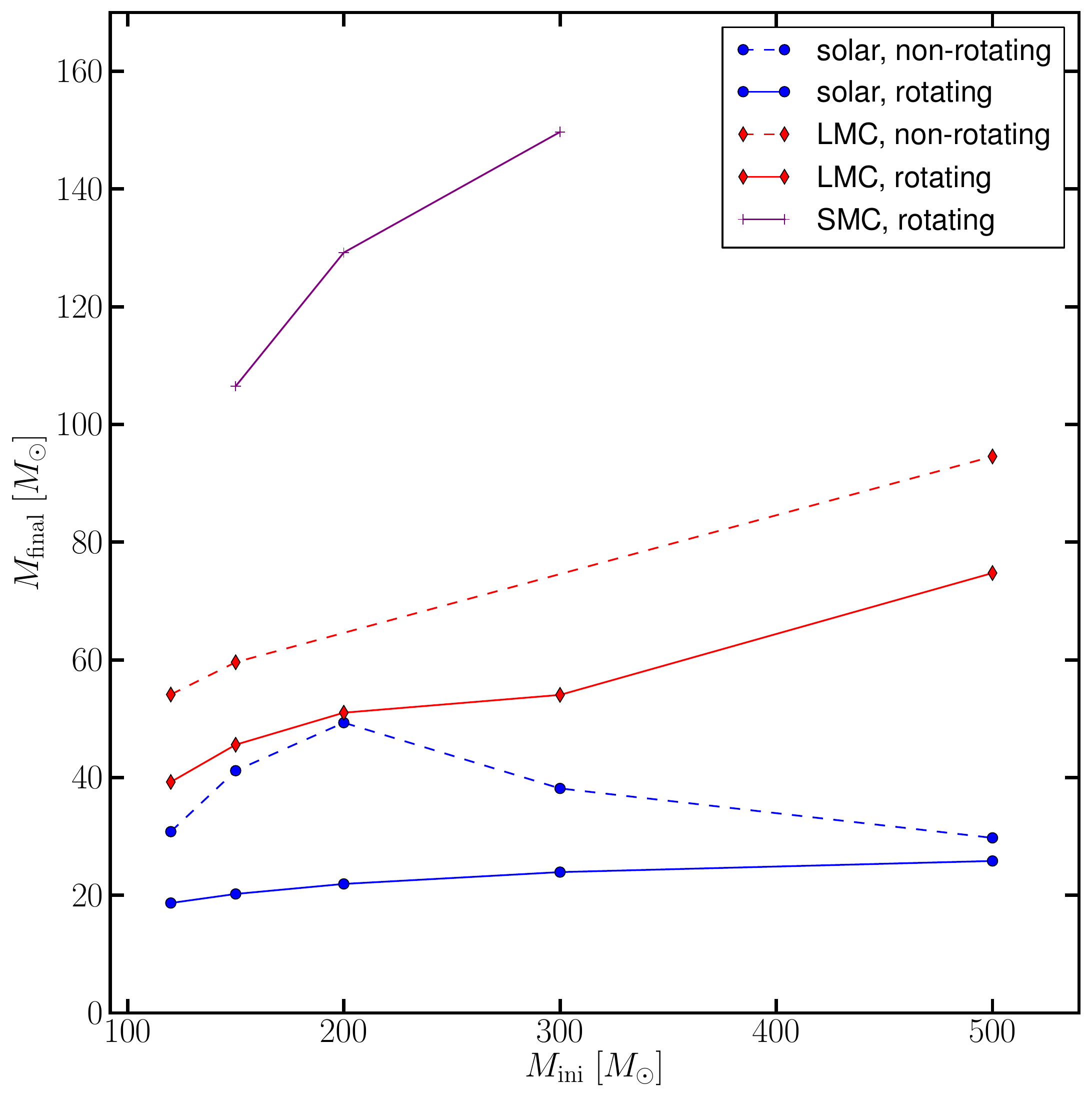}
\caption{Final mass versus initial mass for all our rotating (solid lines) and non-rotating (dashed line) models.} \label{fig:mfinal}
\end{figure}

\begin{figure}
\includegraphics[width=0.5\textwidth]{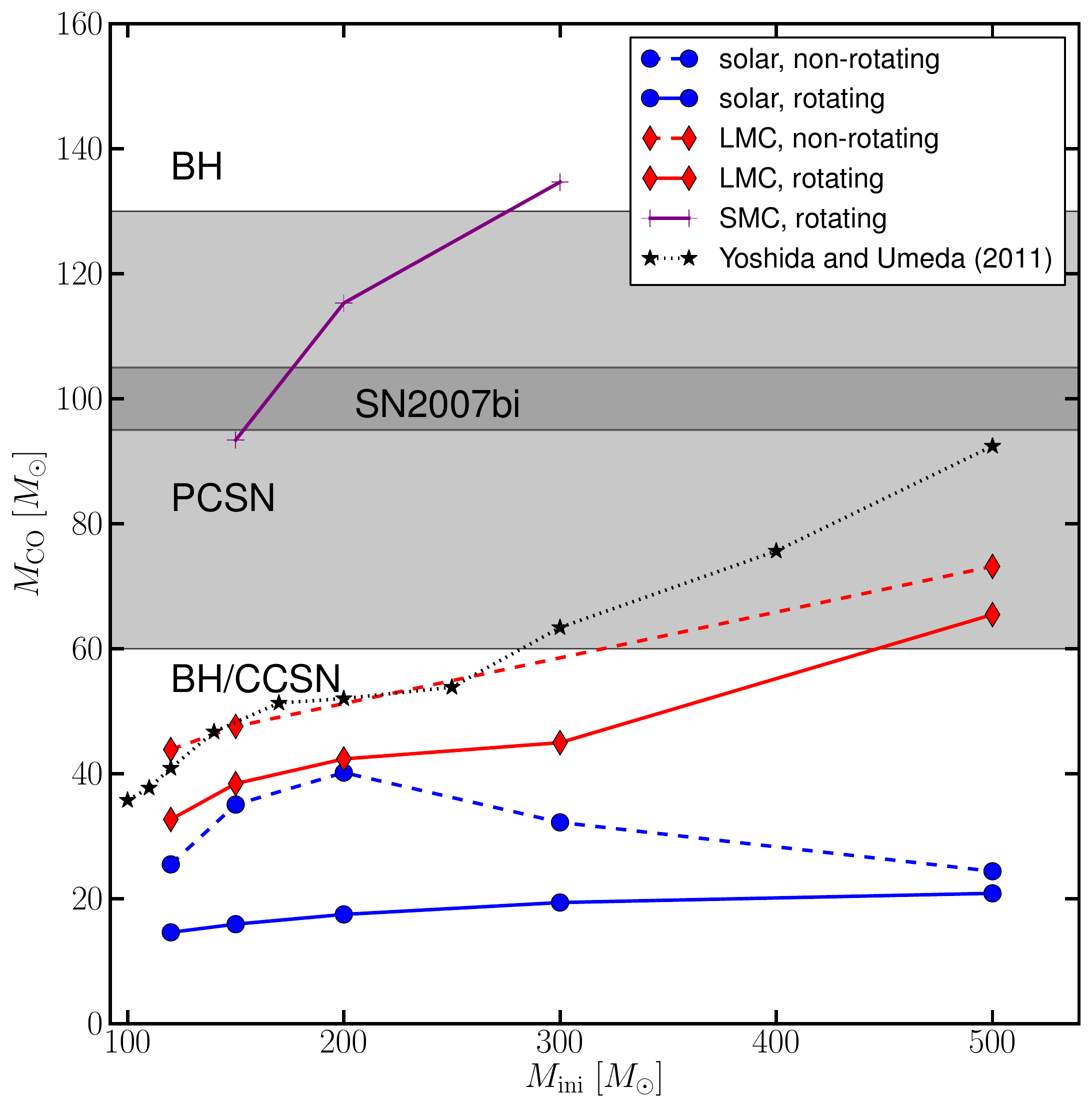}
\caption{Mass of carbon-oxygen core of all the models as a function of the initial mass. The light grey shaded area represents the range of $M_{CO}$, for which the estimated fate is a PCSN. The thin dark grey shaded area corresponds to the estimated $M_{CO}$ of the progenitor of SN2007bi assuming it is a PCSN (see text for more details). The points linked by the dotted black line are from the models of \citet{YH11} at $Z=0.004$, case A.  } \label{fig:mco}
\end{figure}

Fig. \ref{fig:mco} shows how the CO core masses vary as a function of the initial mass, rotation and metallicity.  The CO core ($M_{\textrm{CO}}$) is here defined as the core mass for which the mass fraction of C+O is greater than 75\%.
Since the CO core mass is so close to the total mass, the behavior is the same as for the total mass and for the same reasons. For the rotating solar metallicity models, mass loss is so strong that all models end with roughly the same CO core mass around 20 $M_\odot$. As the metallicity decreases, so does mass loss and thus the LMC and SMC models have higher final CO core masses and the CO core mass does depend on the initial mass in a monotonous way. Finally, non-rotating models lose less mass than their rotating counterpart since they enter the WR phase later and also have less hot surface.

We see in Fig. \ref{fig:mco} that the final CO core mass of our non-rotating models at $Z=0.006$ are slightly smaller than those obtained for non-rotating stars at $Z=0.004$ by \citet{YH11}. We have also compared the evolutionary tracks in the HRD of our non-rotating LMC models and the models of \citet{YH11} case A and found them to be qualitatively very similar. 

As discussed above, the core masses, especially the CO core masses, can be used to {estimate whether or not our models produce PCSN} by using the results of previous studies, which follow the explosion of such massive cores and knowing that VMS with the same CO core masses have similar core evolution from carbon burning onwards. \citet{HEGER02} calculated a grid of models and found that stars with helium cores ($M_{\alpha}$) between 64 and 133 $M_\odot$ produce Pair Creation SuperNova (PCSN) supernova and that stars with more massive $M_{\alpha}$ will collapse to a BH without explosion, confirming the results of previous studies, such as \citet{BAC84}. {The independent results of \citet{CE12} also confirm the CO core mass range that produce PCSNe.}

Let us recall here that
PCSNe occur when very massive stars (VMS) experience an instability in
their core during the neon/oxygen burning stage due to the creation of electron-positron
pairs out of two photons. The creation of pairs in their oxygen-rich core softens the equation of
state, leading to further contraction. This runaway collapse
is predicted to produce a very powerful explosion, in excess of $10^{53}$ erg, disrupting the entire star and leaving no remnant
\citep{BAC84, FCL01}.

 \citet{HEGER02} also find that stars with $M_{\alpha}$ between roughly 40 and 63 $M_\odot$ will undergo violent pulsations induced by the pair-instability leading to strong mass loss but which will not be sufficient to disrupt the core. Thus these stars will eventually undergo core collapse as lower mass stars. Since in our models, the CO core masses are very close to $M_{\alpha}$ (equal to the final total mass in our models, see Table \ref{table:fate}), in this study we assume that our models will produce a PCSN if 60 $M_\odot\leq M_{CO}\leq$ 130 $M_\odot$. In Fig. \ref{fig:mco}, the light grey shaded region corresponds to the zone where one would expect a PCSN, the dark shaded region show the
estimated range of the carbon oxygen core of the progenitor of SN2007bi, as discussed below.

We see in Fig. \ref{fig:mco} that at solar metallicity none of our models is expected to explode as a PCSN. At the metallicity of the LMC, only stars with initial masses above 450 for the rotating models and
above about 300\,$M_\odot$ for the non-rotating case are expected to explode as a PCSN.
At the SMC metallicity, the mass range for the PCSN progenitors is much more favorable.
Extrapolating the points obtained from our models we obtain that all stars in the mass range
between about 100\,$M_\odot$ and 290\,$M_\odot$ could produce PCSNe. Thus our models provide support for the occurrence of PCSNe in the nearby (not so metal poor) universe.

As mentioned earlier, the evolution of a subset of models (the SMC rotating 150, 200 and 300\,$M_\odot$ models and the LMC rotating and non-rotating 500\,$M_\odot$ models) has been followed from the end of core helium burning through to explosions with the Kepler code. The KEPLER simulations confirm that the SMC rotating 150 and 200\,$M_\odot$ models and the rotating and non-rotating 500\,$M_\odot$ LMC models indeed end as a PCSN and their properties will be presented in a forthcoming paper (Whalen et al., in preparation).

Table \ref{table:fate} presents for each of the models, the initial mass ($M_{\rm ini}$), the amount of helium left in the star at the end of the calculation ($M_{\textrm{He}}^{\rm env}$),  and final total mass as well as the predicted fate in terms of the explosion type: PCSN or core-collapse supernova and black hole formation with or without mass ejection (CCSN/BH). The helium core mass ($M_\alpha$) is not given since it is always equal to the final total mass, all our models having lost the entire hydrogen-rich layers.

\subsection{Supernova types produced by VMS and comparison to observed superluminous SNe}

Let us recall that, in VMS, convective cores are very large.
It is larger than 90\% above 200 $M_\odot$ at the start of the evolution and even though it decreases slightly during the evolution, at the end of core H-burning, the convective core occupies more than half of the initial mass in non-rotating models and most of the star in rotating models. This has an important implication concerning the type of supernovae that these VMS will produce. Indeed, even if mass loss is not very strong in SMC models, all the models we have calculated have lost the entire hydrogen rich layers long before the end of helium burning. 
Thus our models predict that all VMS stars in the metallicity range studied will produce either a type Ib or type Ic SN but no type II. 
{Since SN2006gy is a type IIn supernova \citep{SN07}, our models support the idea that the IIn type is due to interaction with circumstellar material rather than by a PCSN from a VMS that had retained its H-rich envelope as discussed in \citet{SN07}.}

{Our models clearly predict no type II SNe but it is not so clear whether the VMS stars that we modeled will produce type Ib or Ic SNe.}
The distinctive feature of SNe Ic is the absence of He\,{\sc i} 
lines in their spectra. The absence of lines, however, is not necessarily indicative of 
a complete absence of helium \citep[see e.\.g.][]{DHLW12}. Other factors such as temperature, density in the region where helium is present are important for the strength of lines. Thus from a theoretical point of view, we are left with some freedom to chose the criterion deciding on whether a star will produce a type Ib or type Ic SN.

\begin{figure}
\includegraphics[width=0.5\textwidth,clip=]{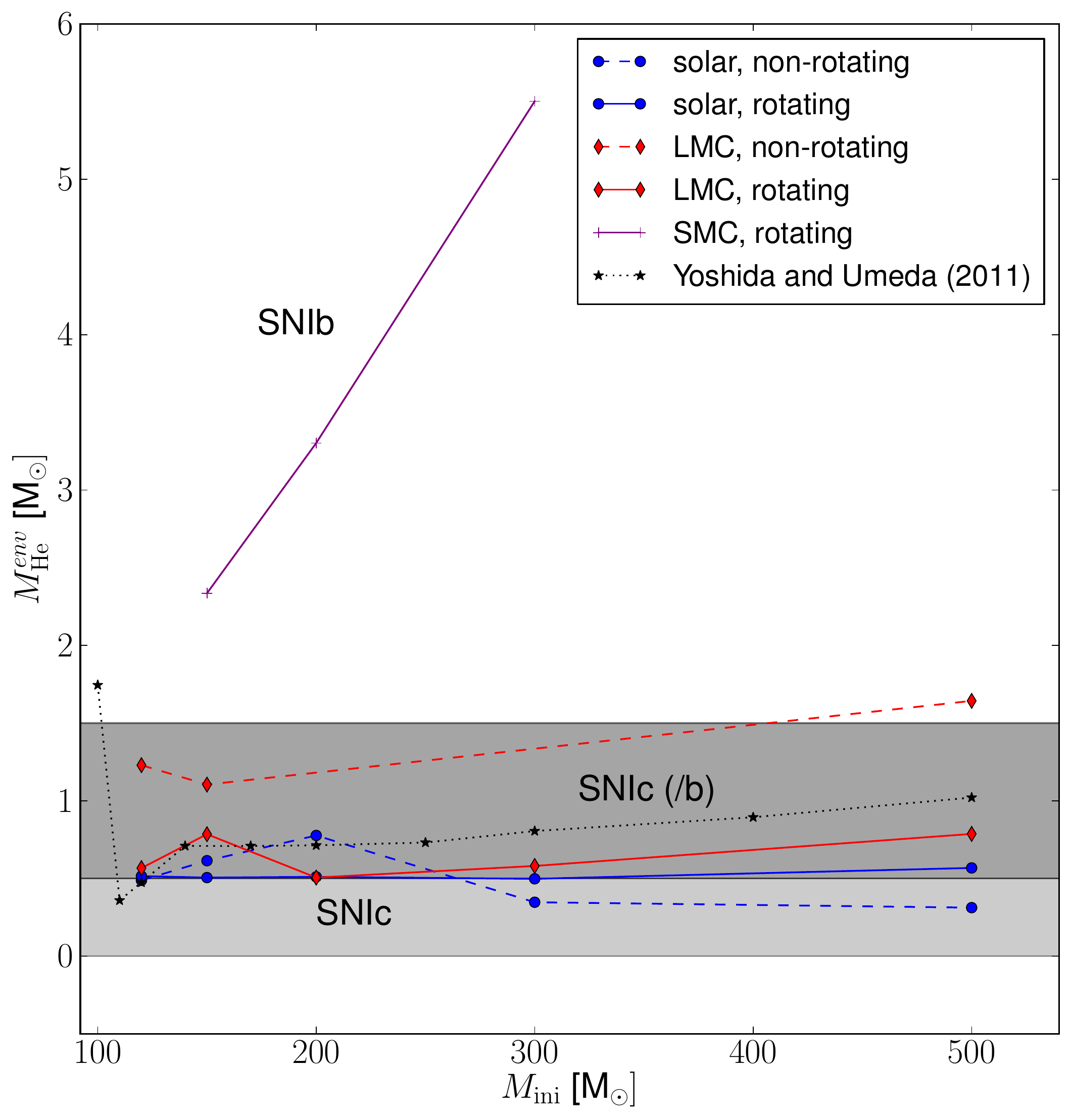}
\includegraphics[width=0.5\textwidth,clip=]{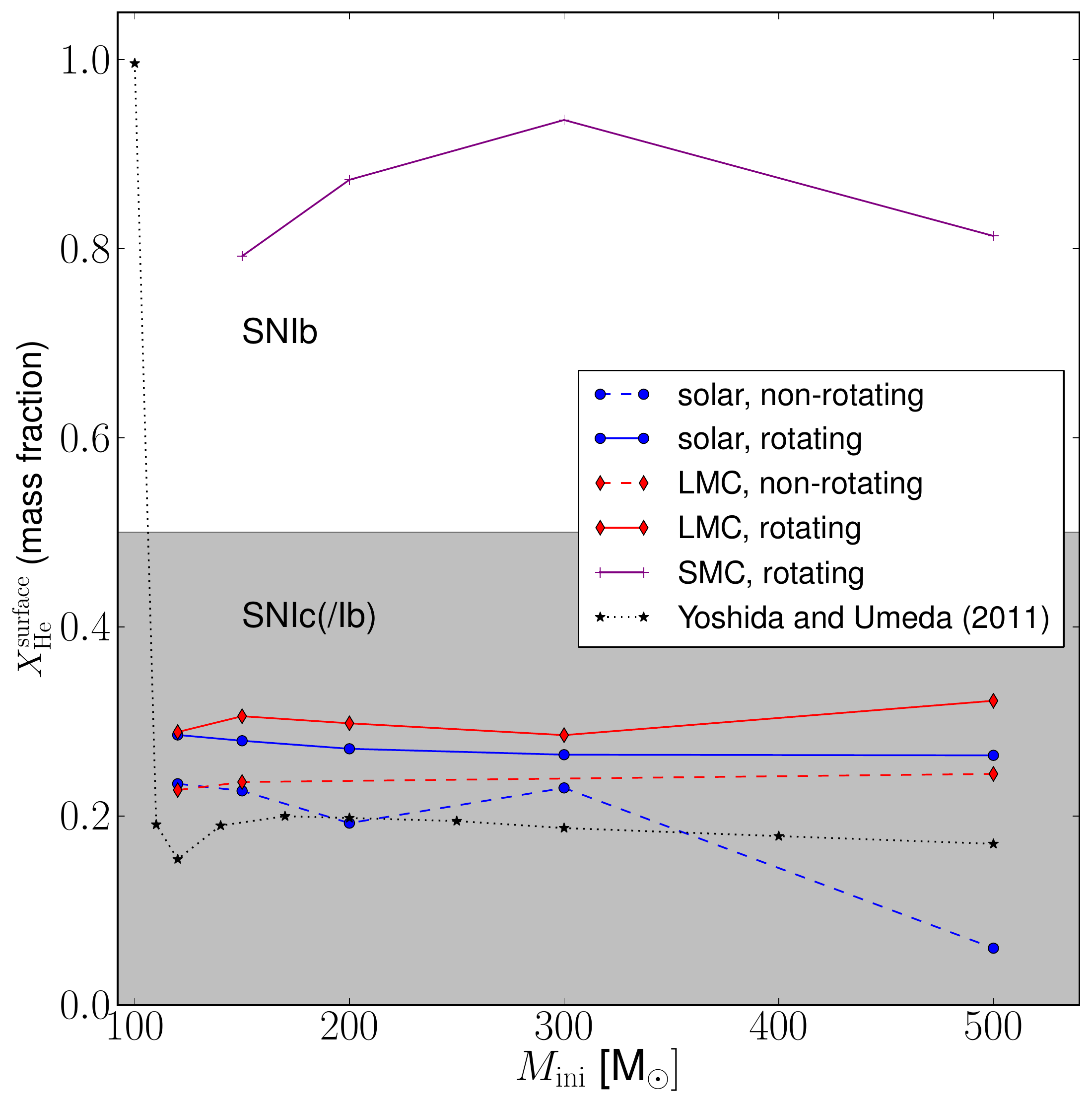}
\caption{Amount of helium left in the star, $M_{\textrm{He}}^{\textrm{env}}$ ({\it top}) and mass fraction of He at the surface, X$^\textrm{surface}_{He}$ ({\it bottom}) at the end of the simulations.} \label{fig:he}
\end{figure}

A first approach is based on the total He mass in the envelope. \citet{WS99} and \citet{Yoon10} choose 0.5 $M_\odot$ while \cite{GC09} proposed 0.6 $M_\odot$ as their limit. However, \cite{GC09} also reported that the choice of He mass limit between 0.6 and 1.5 
$M_\odot$ hardly affects the mass of main sequence ranges for SN Ic/Ib. 
The total mass of helium left in the star at the end of the simulations, $M_{\textrm{He}}^{env}$ for all our models is presented in Fig. \ref{fig:he}. 
Considering that a type Ic supernova is detected only if the total He mass is less than 0.5 $M_\odot$ (light grey area), we see that almost none of our models would predict a type Ic supernova. Only the 300 and 500 non-rotating solar metallicity models would barely qualify. On the other hand, considering that a type Ic supernova is detected if the total He mass is less than 1.5 $M_\odot$ (dark$+$light grey area), then most models at solar and LMC metallicities would produce type Ic while SMC models would produce type Ib supernova.

A second approach is to use the surface He mass fraction, 
X$^\textrm{surface}_{He}$ since only a shell of material is excited at a given time. 
We use the value of X$^\textrm{surface}_{He}=0.5$ as in \cite{YH11} and \cite{Yoon10} as the boundary between type Ib and Ic supernovae (see Fig. \ref{fig:he}, {\it bottom} panel). 
Using this approach, we find that all our LMC and solar metallicity models end as 
SN Ic, whereas those at SMC metallicity end as SNIb. Since it is mostly the SMC models that are predicted to produce PCSN, our models would predict that PCSN are most likely to appear as SNIb. \citet{Yoon10} reported that the He lines are not seen in early- time spectra even though the total He mass is as large as 1.0 $M_\odot$ if He is well mixed with CO thus detailed spectral 
modeling should be used to put tighter constraints on the supernova type.

\subsubsection{Are superluminous SNe PCSNe?}

We have evaluated the initial mass range that might produce PCSN using the final CO core mass of our models. We can now see how our models compare with observed superluminous SNe (SLSNe). Since none of our models retain hydrogen, we will not try to determine a possible initial mass for the progenitor of SNIIn SN2006gy. Interactions with a dense
circumstellar medium is a possible scenario for such SLSNe \citep{GALYAM12}.
The case of SN2007bi is more interesting since this SLSN is a type Ic supernova and its light curve is easily explained by a PCSN \citep{GALYAM09}. The light curve of this supernova can be explained by a PCSN with a CO core mass equal to roughly 100 $M_\odot$.
\citet{YH11} calculated models at $Z=0.004$ (the estimated metallicity for SN2007bi) and find that models with an initial mass of $\sim$500 $M_\odot$ end with the required CO core
using a standard mass loss prescription (case A) as shown in Fig. \ref{fig:mco}. Assuming that SN2007bi had a SMC metallicity (which is still within uncertainties), we see in  Fig. \ref{fig:mco} that the desired CO core mass can be obtained from a much lower initial mass range, roughly between 160 and 175 $M_\odot$, which makes the probability of such events much higher. 

There are, however, several issues with the PCSN scenario to explain the properties of SN2007bi. First, all stellar evolution models retain some helium at their surface and it is not clear whether these models would be observed as type Ic SNe. Second, the synthetic spectra from PCSNe seem to be much redder than the observed spectrum of SN2007bi and other SLSNe \citep{DHW12,DWL13}. Other possible scenarios proposed for SLSNe are energetic SNe \citep[see e.\,g.][]{YH11} and magnetar-driven explosions \citep[see][ and references therein]{DHW12}.

\subsection{GRBs from VMS?}\label{grb}
\citet{YDL12} calculated a grid of zero-metallicity rotating stars, including the Taylor-Spruit dynamo for the interaction between rotation and magnetic fields. They find that fast rotating stars with an initial mass below about 200 $M_\odot$ retain enough angular momentum in their cores in order to produce a collapsar \citep[$j>j_{\textrm Kerr, lso}$][]{W93} or a magnetar \citep[see e.\,g.][]{W00,BDL07,DOO12}.

\begin{figure*}
\includegraphics[width=0.3\textwidth,clip=]{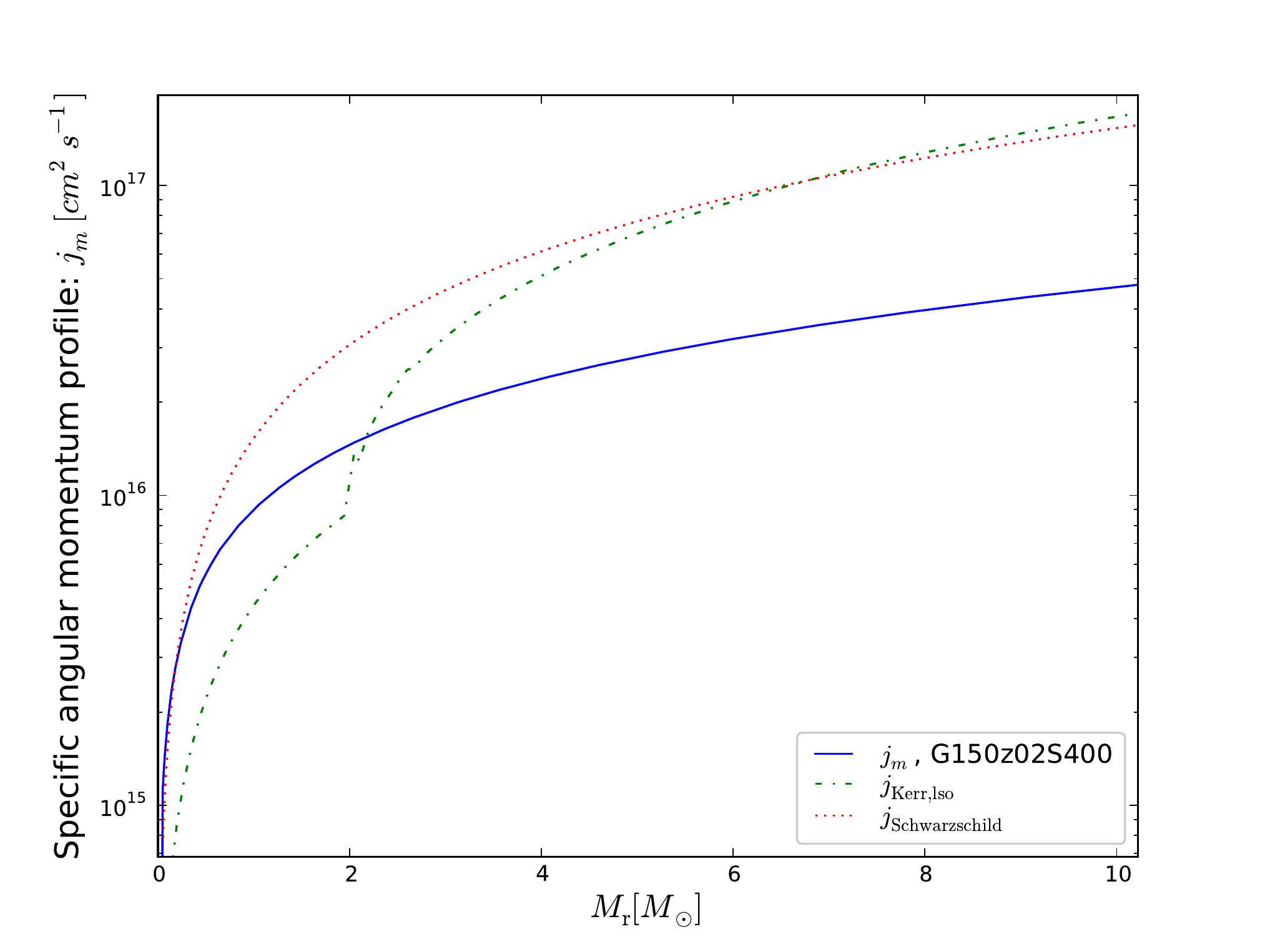}\includegraphics[width=0.3\textwidth,clip=]{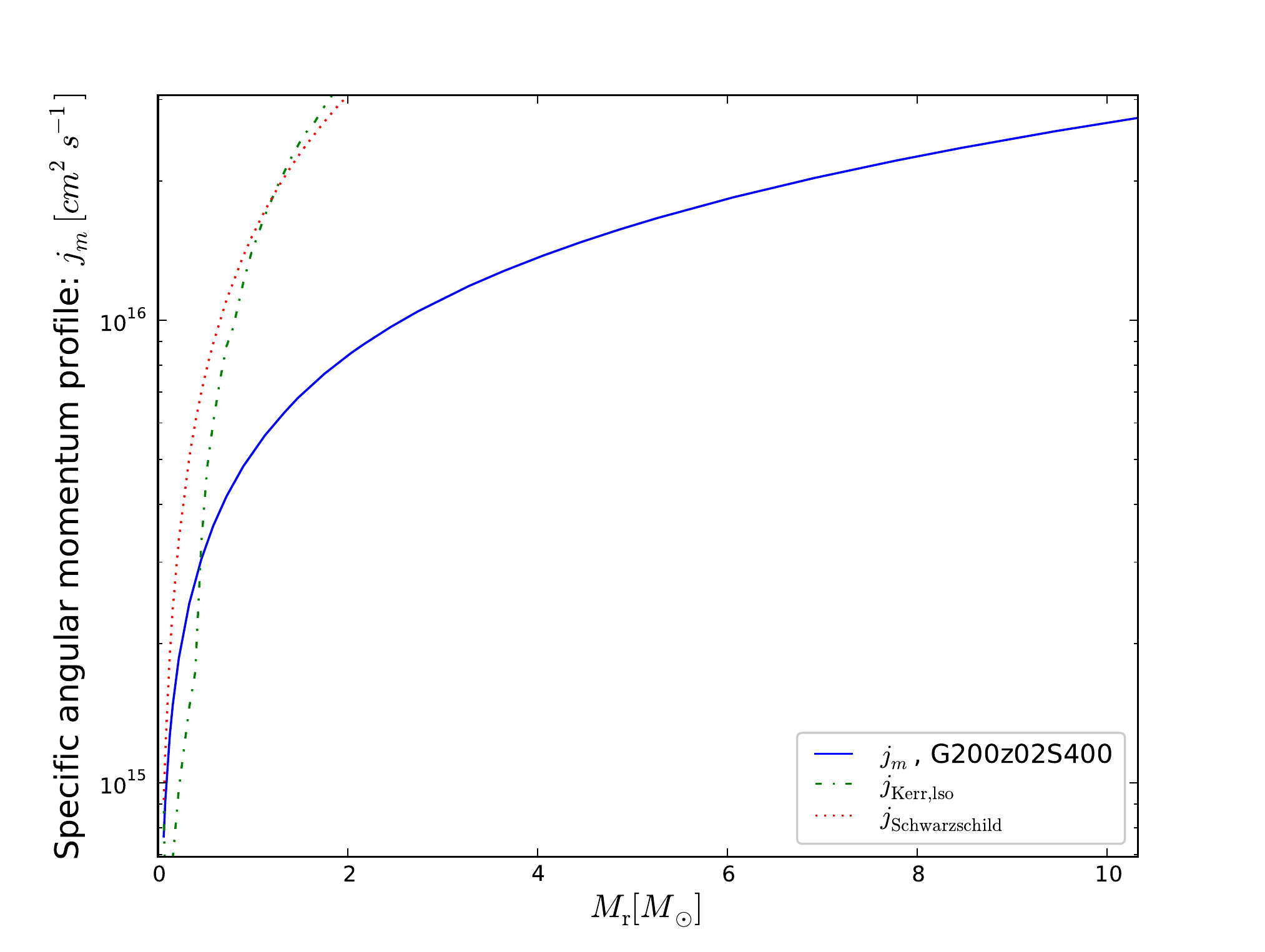}\includegraphics[width=0.3\textwidth,clip=]{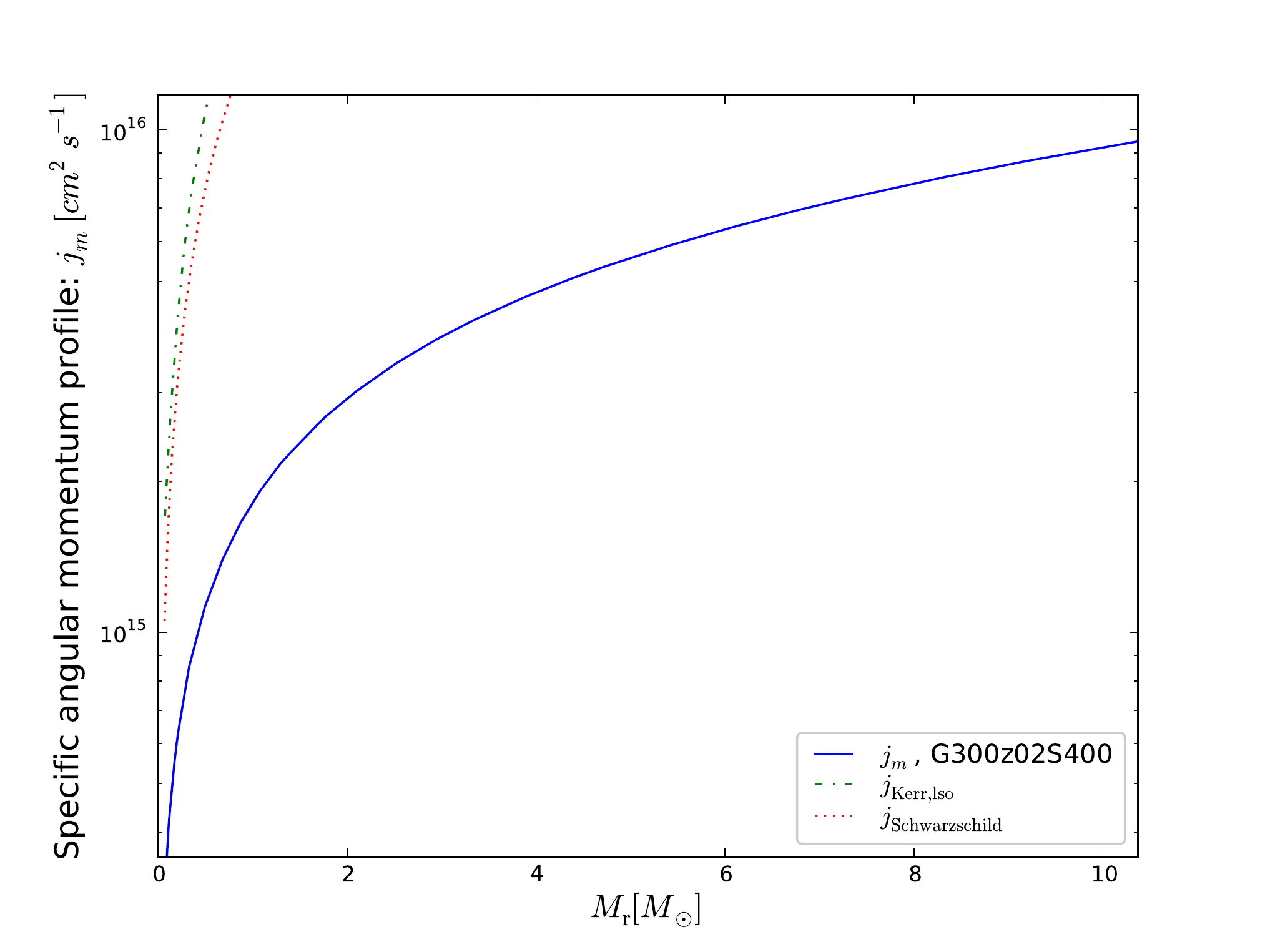}
\caption{Specific angular momentum profile, $j_\mathrm{m}$, as a function of the Lagrangian mass coordinate in the core of the SMC rotating 150, 200, 300 $M_\odot$ models, plotted at the end of the calculations (solid line). The dash-dotted line is $j_{\rm Kerr,lso}=r_{\rm LSO}\,c$ \citep[ p. 428]{ST83}, where the radius of the last stable orbit, $r_{\rm LSO}$, is given by $r_{\rm ms}$ in formula (12.7.24) from \citet[p. 362]{ST83} for circular orbit in the Kerr metric. $j_{\rm Kerr,lso}$ is the minimum specific angular momentum necessary to form an accretion disc around a rotating black hole. $j_\mathrm{Schwarzschild}=\sqrt{12}Gm/c$ (dotted line) is the minimum specific angular momentum necessary for a non-rotating black hole, for reference.} \label{fig:grb}
\end{figure*}

The evolution of the surface velocity was described in Sect. \ref{vsurf}. Only models at SMC retain a significant amount of rotation
during their evolution {(see angular momentum contained in the CO core at the end of helium burning in the last column of
Table \ref{Table:endHe})} but do they retain enough angular momentum for rotation to affect the fate of the star? The angular
momentum profile of the SMC models is presented in Fig. \ref{fig:grb}. Note that our models do not include the Taylor-Spruit dynamo
so represent the most optimistic (highest possible) prediction concerning the angular momentum in the core of these models. Mass loss
in the 300\,$M_\odot$ model is too strong for the core to retain enough angular momentum for rotation to impact the death of this
model. In the 200\,$M_\odot$ model, and even more so in the 150\,$M_\odot$ model, however, the central part of the core retain a
significant amount of angular momentum that could potentially affect the death of the star. {Since the role of rotation is
very modest from carbon until just after the end of core silicon burning, even for extremely fast rotators \citep[see
e.\,g.][]{HMM05,CL13}, we do not expect rotation to affect significantly the fate of stars that are predicted to explode as PCSN
during neon-oxygen burning. However, as discussed in \citet[][and references therein]{YDL12}, the large angular momentum content is
most interesting for the stars that just fall short of the minimum CO core mass for PCSN \citep[since fast rotation plays an
important role during the early collapse][]{OTT04,CON11,CL13}.} Indeed, without rotation, these stars would produce a BH following a possible pulsation pair-creation phase, whereas with rotation, these stars could produce energetic asymmetric explosions (GRBs or magnetars). Since the 150 $M_\odot$ model is predicted to explode as a PCSN, we thus do not expect the models presented in this grid to produce GRBs or magnetars but such energetic asymmetric explosions are likely to take place in lower mass and lower metallicity stars \citep[see][]{HMM05,YL05,WH06}.

\section{Summary and Conclusion}

We have calculated a grids of stellar models of very massive stars at SMC, 
LMC and solar metallicities. Our study is motivated by the finding of very 
massive stars including R136a1 \citep{PAC10} and the observation of 
PCSN candidate, SN 2007bi by \citep{GALYAM09}.

The main results of this study are the following:
\begin{itemize}
\item VMS possess very large convective cores during the MS phase. Typically, in a 200 $M_\odot$ model on the ZAMS
the convective core extends  over more than 90\% of the total mass.
\item Even in models with no rotation, due to the importance of the convective core, VMS stars evolve nearly homogeneously.
\item Most of the very massive stars (all at solar $Z$) remain in blue regions of the HR diagram and do not go through a luminous blue variable phase.
\item They all enter into the WR phase and their typical evolution will be Of - WNL-  WNE - WC/WO.
\item Due to increasing mass loss rates with the mass, very different initial mass stars end with similar final masses. As a consequence
very different initial masses may during some of their evolutionary phases occupy very similar positions in the HRD.
\item A significant proportion of the total stellar lifetimes of VMS is spent in the WR phase.
At solar metallicity between 16 and 43\% depending on the initial mass and rotation.
These proportions decrease with the metallicity to values between 12 and 39\% for the LMC metallicity.
\item A WC star with high Ne ($^{20}$Ne) and Mg  ($^{24}$Mg) abundances at the surface has necessarily a VMS as progenitor.
\item At solar metallicity none of our model is expected to explode as a PCSN. 
At the metallicity of the LMC, only stars with initial masses above 450 for the rotating models and
above about 300 $M_\odot$ for the non-rotating case are expected to explode as a PCSN.
At the SMC metallicity, the mass range for the PCSN progenitors is much more favorable.
We obtain that all rotating stars in the mass range
between about 100 $M_\odot$ and 290 $M_\odot$ would produce PCSNe.
\item All the models we have calculated have lost the entire hydrogen rich layers long before the end of helium burning. 
Thus our models predict that all VMS stars in the metallicity range studied will produce either a type Ib or type Ic SN but no type II.
\item Assuming that SN2007bi had a SMC metallicity, we determine an initial mass for the progenitor  between 160 and 175 $M_\odot$.
\item We do not expect that the models presented in this grid produce GRBs or magnetars. The reason for that is that either they lose too much angular momentum by mass loss or they avoid the formation of a neutron star or BH because they explode as PCSN. Lower mass stars at low metallicities ($Z\lesssim 0.002$), however, may retain enough angular momentum as in metal free stars \citep[see][]{YDL12,CE12} for rotation (and magnetic fields) to play a significant role in their explosion.
\end{itemize}

To conclude this paper, we can wonder what the importance of the VMS
on the scale of galaxies is?
Are very massive stars so rare that whatever their evolution, their impact on energy and mass outputs will anyway be very low?
Considering a Salpeter IMF, the number of stars with masses 
between 120 and 500  $M_{\odot}$ corresponds to only about 2\% of the total number of stars with masses between 8 and 500 $M_{\odot}$. 
So they are indeed only very few!
On the other hand, one explosion can release a great amount of energy and mass into the interstellar medium. 
Typically a 200  $M_{\odot}$ star releases about ten times more mass than a 20 $M_{\odot}$ star. 
If we roughly suppose that for hundred 20 $M_{\odot}$ stars there are only two 200 $M_{\odot}$ star, this means that the 200 $M_{\odot}$ stars contribute to the release of mass at a level corresponding to about 20\% of the release of mass by 20 $M_{\odot}$, which is by far not negligible. Of course this is a rough estimate but, as a rule of thumb we can say that any quantity released by a VMS $\sim$ tenfold
intensity compared to that of a typical, 20-Msol star will make a
non-negligible difference in the overall budget of this quantity at the level of a galaxy. 
For instance, the high bolometric
luminosities, stellar temperatures and mass loss rates of VMS imply
that they will contribute significantly to the radiative and mechanical
feedback from stars in high mass clusters at ages prior to the first 
supernovae \citep{PAC10}. Core-collapse SNe produce of the order of $0.05\,M_\odot$ (ejected masses) of iron, $1\,M_\odot$ of each of the $\alpha-$elements. According to the production factors in Table 4 in \citet{HEGER02}, PCSN produce up to $40\,M_\odot$ of iron, of the order of $30\,M_\odot$ of oxygen and silicon and of the order of $5-10\,M_\odot$ of the other $\alpha-$elements. Considering that PCSN may occur up to SMC metallicity and represent 2\% of SNe at a given metallicity, their contribution to the chemical enrichment of galaxies is significant, especially in the case of iron, oxygen and silicon.

\section*{Acknowledgments}
The authors thank A. Heger and D. Whalen for modelling the evolution of a subset of our models with the KEPLER code and fruitful discussions. They also thank T. Yoshida for supplying the final mass data from the \citet{YH11} paper.
N.Yusof acknowledges financial support by the Ministry
of Higher Education and University of Malaya under Higher
Education Academic Training Scheme and the Commonwealth
Scholarship Commission for the Split-Site PhD 2010-2011 program
tenable at Keele University. R. Hirschi acknowledges support from the World Premier International Research
Center Initiative (WPI Initiative), MEXT, Japan and from the Eurogenesis EUROCORE programme. The research leading to these results has received funding from the European Research Council under the European Union's Seventh Framework Programme (FP/2007-2013) / ERC Grant Agreement n. 306901. 
N. Yusof and H.A. Kassim acknowledge support from Fundamental Research Grant Scheme (FRGS/FP009-2012A) under Ministry of Higher Education of Malaysia.

\bibliographystyle{mn2e}
\bibliography{grid}


\label{lastpage}

\end{document}